\documentclass{article}
\usepackage{ieeetrantools}
\usepackage{subcaption}
\usepackage{graphicx}
\usepackage{spconf,amsmath,graphicx}
\usepackage{setspace}
\usepackage{tabularx,booktabs}
\usepackage{dingbat}
\usepackage{diagbox}
\usepackage{float}
\usepackage{multirow} 
\usepackage[hyphens]{url}
\usepackage[pagebackref,breaklinks,colorlinks]{hyperref}
\usepackage{xcolor}
\usepackage{colortbl}
\usepackage{amsfonts, amsmath, amsthm, amssymb}
\usepackage{subcaption}
\hypersetup{breaklinks=true}
\usepackage{tikz}
\usepackage{textcomp}
\usepackage{lipsum}
\usepackage{svg}
\usepackage{enumitem}
\usepackage[nameinlink]{cleveref}
\usepackage{amssymb}
\usepackage{pifont}
\usepackage{scalerel}
\usepackage{stackengine}
\usepackage{listings}
\setlist[itemize]{noitemsep}

\newcommand\blfootnote[1]{%
  \begingroup
  \renewcommand\thefootnote{}\footnote{#1}%
  \addtocounter{footnote}{-1}%
  \endgroup
}
\lstdefinelanguage{PowerShell}{
	morekeywords={
		Add-Content,Add-PSSnapin,Clear-Content,Clear-History,Clear-Host,Clear-Item,Clear-ItemProperty,Clear-Variable,Compare-Object,Connect-PSSession,ConvertFrom-String,Convert-Path,Copy-Item,Copy-ItemProperty,Disable-PSBreakpoint,Disconnect-PSSession,Enable-PSBreakpoint,Enter-PSSession,Exit-PSSession,Export-Alias,Export-Csv,Export-PSSession,ForEach-Object,Format-Custom,Format-Hex,Format-List,Format-Table,Format-Wide,Get-Alias,Get-ChildItem,Get-Clipboard,Get-Command,Get-ComputerInfo,Get-Content,Get-History,Get-Item,Get-ItemProperty,Get-ItemPropertyValue,Get-Job,Get-Location,Get-Member,Get-Module,Get-Process,Get-PSBreakpoint,Get-PSCallStack,Get-PSDrive,Get-PSSession,Get-PSSnapin,Get-Service,Get-TimeZone,Get-Unique,Get-Variable,Get-WmiObject,Group-Object,help,Import-Alias,Import-Csv,Import-Module,Import-PSSession,Invoke-Command,Invoke-Expression,Invoke-History,Invoke-Item,Invoke-RestMethod,Invoke-WebRequest,Invoke-WmiMethod,Measure-Object,mkdir,Move-Item,Move-ItemProperty,New-Alias,New-Item,New-Module,New-PSDrive,New-PSSession,New-PSSessionConfigurationFile,New-Variable,Out-GridView,Out-Host,Out-Printer,Pop-Location,powershell_ise.exe,Push-Location,Receive-Job,Receive-PSSession,Remove-Item,Remove-ItemProperty,Remove-Job,Remove-Module,Remove-PSBreakpoint,Remove-PSDrive,Remove-PSSession,Remove-PSSnapin,Remove-Variable,Remove-WmiObject,Rename-Item,Rename-ItemProperty,Resolve-Path,Resume-Job,Select-Object,Select-String,Set-Alias,Set-Clipboard,Set-Content,Set-Item,Set-ItemProperty,Set-Location,Set-PSBreakpoint,Set-TimeZone,Set-Variable,Set-WmiInstance,Show-Command,Sort-Object,Start-Job,Start-Process,Start-Service,Start-Sleep,Stop-Job,Stop-Process,Stop-Service,Suspend-Job,Tee-Object,Trace-Command,Wait-Job,Where-Object,Write-Output
	},
	morekeywords={
		Add-AppxPackage,Add-AppxProvisionedPackage,Add-AppxVolume,Add-BitsFile,Add-CertificateEnrollmentPolicyServer,Add-Computer,Add-Content,Add-History,Add-JobTrigger,Add-KdsRootKey,Add-LocalGroupMember,Add-Member,Add-PSSnapin,Add-Type,Add-WindowsCapability,Add-WindowsDriver,Add-WindowsImage,Add-WindowsPackage,Checkpoint-Computer,Clear-Content,Clear-EventLog,Clear-History,Clear-Item,Clear-ItemProperty,Clear-KdsCache,Clear-RecycleBin,Clear-Tpm,Clear-Variable,Clear-WindowsCorruptMountPoint,Compare-Object,Complete-BitsTransfer,Complete-DtiagnosticTransaction,Complete-Transaction,Confirm-SecureBootUEFI,Connect-PSSession,Connect-WSMan,ConvertFrom-Csv,ConvertFrom-Json,ConvertFrom-SecureString,ConvertFrom-String,ConvertFrom-StringData,Convert-Path,Convert-String,ConvertTo-Csv,ConvertTo-Html,ConvertTo-Json,ConvertTo-ProcessMitigationPolicy,ConvertTo-SecureString,ConvertTo-TpmOwnerAuth,ConvertTo-Xml,Copy-Item,Copy-ItemProperty,Debug-Job,Debug-Process,Debug-Runspace,Disable-AppBackgroundTaskDiagnosticLog,Disable-ComputerRestore,Disable-JobTrigger,Disable-LocalUser,Disable-PSBreakpoint,Disable-PSRemoting,Disable-PSSessionConfiguration,Disable-RunspaceDebug,Disable-ScheduledJob,Disable-TlsCipherSuite,Disable-TlsEccCurve,Disable-TlsSessionTicketKey,Disable-TpmAutoProvisioning,Disable-WindowsErrorReporting,Disable-WindowsOptionalFeature,Disable-WSManCredSSP,Disconnect-PSSession,Disconnect-WSMan,Dismount-AppxVolume,Dismount-WindowsImage,Enable-AppBackgroundTaskDiagnosticLog,Enable-ComputerRestore,Enable-JobTrigger,Enable-LocalUser,Enable-PSBreakpoint,Enable-PSRemoting,Enable-PSSessionConfiguration,Enable-RunspaceDebug,Enable-ScheduledJob,Enable-TlsCipherSuite,Enable-TlsEccCurve,Enable-TlsSessionTicketKey,Enable-TpmAutoProvisioning,Enable-WindowsErrorReporting,Enable-WindowsOptionalFeature,Enable-WSManCredSSP,Enter-PSHostProcess,Enter-PSSession,Exit-PSHostProcess,Exit-PSSession,Expand-WindowsCustomDataImage,Expand-WindowsImage,Export-Alias,Export-BinaryMiLog,Export-Certificate,Export-Clixml,Export-Console,Export-Counter,Export-Csv,Export-FormatData,Export-ModuleMember,Export-PfxCertificate,Export-ProvisioningPackage,Export-PSSession,Export-StartLayout,Export-StartLayoutEdgeAssets,Export-TlsSessionTicketKey,Export-Trace,Export-WindowsCapabilitySource,Export-WindowsDriver,Export-WindowsImage,Find-Package,Find-PackageProvider,ForEach-Object,Format-Custom,Format-List,Format-SecureBootUEFI,Format-Table,Format-Wide,Get-Acl,Get-Alias,Get-AppxDefaultVolume,Get-AppxPackage,Get-AppxPackageManifest,Get-AppxProvisionedPackage,Get-AppxVolume,Get-AuthenticodeSignature,Get-BitsTransfer,Get-Certificate,Get-CertificateAutoEnrollmentPolicy,Get-CertificateEnrollmentPolicyServer,Get-CertificateNotificationTask,Get-ChildItem,Get-CimAssociatedInstance,Get-CimClass,Get-CimInstance,Get-CimSession,Get-Clipboard,Get-CmsMessage,Get-Command,Get-ComputerInfo,Get-ComputerRestorePoint,Get-Content,Get-ControlPanelItem,Get-Counter,Get-Credential,Get-Culture,Get-DAPolicyChange,Get-Date,Get-DeliveryOptimizationLog,Get-DeliveryOptimizationPerfSnap,Get-DeliveryOptimizationPerfSnapThisMonth,Get-DeliveryOptimizationStatus,Get-DODownloadMode,Get-DOPercentageMaxBackgroundBandwidth,Get-DOPercentageMaxForegroundBandwidth,Get-Event,Get-EventLog,Get-EventSubscriber,Get-ExecutionPolicy,Get-FormatData,Get-Help,Get-History,Get-Host,Get-HotFix,Get-Item,Get-ItemProperty,Get-ItemPropertyValue,Get-Job,Get-JobTrigger,Get-KdsConfiguration,Get-KdsRootKey,Get-LocalGroup,Get-LocalGroupMember,Get-LocalUser,Get-Location,Get-Member,Get-Module,Get-Package,Get-PackageProvider,Get-PackageSource,Get-PfxCertificate,Get-PfxData,Get-PmemDisk,Get-PmemPhysicalDevice,Get-PmemUnusedRegion,Get-Process,Get-ProcessMitigation,Get-ProvisioningPackage,Get-PSBreakpoint,Get-PSCallStack,Get-PSDrive,Get-PSHostProcessInfo,Get-PSProvider,Get-PSReadlineKeyHandler,Get-PSReadlineOption,Get-PSSession,Get-PSSessionCapability,Get-PSSessionConfiguration,Get-PSSnapin,Get-Random,Get-Runspace,Get-RunspaceDebug,Get-ScheduledJob,Get-ScheduledJobOption,Get-SecureBootPolicy,Get-SecureBootUEFI,Get-Service,Get-TimeZone,Get-TlsCipherSuite,Get-TlsEccCurve,Get-Tpm,Get-TpmEndorsementKeyInfo,Get-TpmSupportedFeature,Get-TraceSource,Get-Transaction,Get-TroubleshootingPack,Get-TrustedProvisioningCertificate,Get-TypeData,Get-UICulture,Get-Unique,Get-Variable,Get-WIMBootEntry,Get-WinAcceptLanguageFromLanguageListOptOut,Get-WinCultureFromLanguageListOptOut,Get-WinDefaultInputMethodOverride,Get-WindowsCapability,Get-WindowsDeveloperLicense,Get-WindowsDriver,Get-WindowsEdition,Get-WindowsErrorReporting,Get-WindowsImage,Get-WindowsImageContent,Get-WindowsOptionalFeature,Get-WindowsPackage,Get-WindowsSearchSetting,Get-WinEvent,Get-WinHomeLocation,Get-WinLanguageBarOption,Get-WinSystemLocale,Get-WinUILanguageOverride,Get-WinUserLanguageList,Get-WmiObject,Get-WSManCredSSP,Get-WSManInstance,Group-Object,Import-Alias,Import-BinaryMiLog,Import-Certificate,Import-Clixml,Import-Counter,Import-Csv,Import-LocalizedData,Import-Module,Import-PackageProvider,Import-PfxCertificate,Import-PSSession,Import-StartLayout,Import-TpmOwnerAuth,Initialize-PmemPhysicalDevice,Initialize-Tpm,Install-Package,Install-PackageProvider,Install-ProvisioningPackage,Install-TrustedProvisioningCertificate,Invoke-CimMethod,Invoke-Command,Invoke-CommandInDesktopPackage,Invoke-DscResource,Invoke-Expression,Invoke-History,Invoke-Item,Invoke-RestMethod,Invoke-TroubleshootingPack,Invoke-WebRequest,Invoke-WmiMethod,Invoke-WSManAction,Join-DtiagnosticResourceManager,Join-Path,Limit-EventLog,Measure-Command,Measure-Object,Mount-AppxVolume,Mount-WindowsImage,Move-AppxPackage,Move-Item,Move-ItemProperty,New-Alias,New-CertificateNotificationTask,New-CimInstance,New-CimSession,New-CimSessionOption,New-DtiagnosticTransaction,New-Event,New-EventLog,New-FileCatalog,New-Item,New-ItemProperty,New-JobTrigger,New-LocalGroup,New-LocalUser,New-Module,New-ModuleManifest,New-NetIPsecAuthProposal,New-NetIPsecMainModeCryptoProposal,New-NetIPsecQuickModeCryptoProposal,New-Object,New-PmemDisk,New-ProvisioningRepro,New-PSDrive,New-PSRoleCapabilityFile,New-PSSession,New-PSSessionConfigurationFile,New-PSSessionOption,New-PSTransportOption,New-PSWorkflowExecutionOption,New-ScheduledJobOption,New-SelfSignedCertificate,New-Service,New-TimeSpan,New-TlsSessionTicketKey,New-Variable,New-WebServiceProxy,New-WindowsCustomImage,New-WindowsImage,New-WinEvent,New-WinUserLanguageList,New-WSManInstance,New-WSManSessionOption,Optimize-AppxProvisionedPackages,Optimize-WindowsImage,Out-Default,Out-File,Out-GridView,Out-Host,Out-Null,Out-Printer,Out-String,Pop-Location,Protect-CmsMessage,Publish-DscConfiguration,Push-Location,Read-Host,Receive-DtiagnosticTransaction,Receive-Job,Receive-PSSession,Register-ArgumentCompleter,Register-CimIndicationEvent,Register-EngineEvent,Register-ObjectEvent,Register-PackageSource,Register-PSSessionConfiguration,Register-ScheduledJob,Register-WmiEvent,Remove-AppxPackage,Remove-AppxProvisionedPackage,Remove-AppxVolume,Remove-BitsTransfer,Remove-CertificateEnrollmentPolicyServer,Remove-CertificateNotificationTask,Remove-CimInstance,Remove-CimSession,Remove-Computer,Remove-Event,Remove-EventLog,Remove-Item,Remove-ItemProperty,Remove-Job,Remove-JobTrigger,Remove-LocalGroup,Remove-LocalGroupMember,Remove-LocalUser,Remove-Module,Remove-PmemDisk,Remove-PSBreakpoint,Remove-PSDrive,Remove-PSReadlineKeyHandler,Remove-PSSession,Remove-PSSnapin,Remove-TypeData,Remove-Variable,Remove-WindowsCapability,Remove-WindowsDriver,Remove-WindowsImage,Remove-WindowsPackage,Remove-WmiObject,Remove-WSManInstance,Rename-Computer,Rename-Item,Rename-ItemProperty,Rename-LocalGroup,Rename-LocalUser,Repair-WindowsImage,Reset-ComputerMachinePassword,Resolve-DnsName,Resolve-Path,Restart-Computer,Restart-Service,Restore-Computer,Resume-BitsTransfer,Resume-Job,Resume-ProvisioningSession,Resume-Service,Save-Help,Save-Package,Save-WindowsImage,Select-Object,Select-String,Select-Xml,Send-DtiagnosticTransaction,Send-MailMessage,Set-Acl,Set-Alias,Set-AppBackgroundTaskResourcePolicy,Set-AppxDefaultVolume,Set-AppXProvisionedDataFile,Set-AuthenticodeSignature,Set-BitsTransfer,Set-CertificateAutoEnrollmentPolicy,Set-CimInstance,Set-Clipboard,Set-Content,Set-Culture,Set-Date,Set-DODownloadMode,Set-DOPercentageMaxBackgroundBandwidth,Set-DOPercentageMaxForegroundBandwidth,Set-DscLocalConfigurationManager,Set-ExecutionPolicy,Set-Item,Set-ItemProperty,Set-JobTrigger,Set-KdsConfiguration,Set-LocalGroup,Set-LocalUser,Set-Location,Set-PackageSource,Set-ProcessMitigation,Set-PSBreakpoint,Set-PSDebug,Set-PSReadlineKeyHandler,Set-PSReadlineOption,Set-PSSessionConfiguration,Set-ScheduledJob,Set-ScheduledJobOption,Set-SecureBootUEFI,Set-Service,Set-StrictMode,Set-TimeZone,Set-TpmOwnerAuth,Set-TraceSource,Set-Variable,Set-WinAcceptLanguageFromLanguageListOptOut,Set-WinCultureFromLanguageListOptOut,Set-WinDefaultInputMethodOverride,Set-WindowsEdition,Set-WindowsProductKey,Set-WindowsSearchSetting,Set-WinHomeLocation,Set-WinLanguageBarOption,Set-WinSystemLocale,Set-WinUILanguageOverride,Set-WinUserLanguageList,Set-WmiInstance,Set-WSManInstance,Set-WSManQuickConfig,Show-Command,Show-ControlPanelItem,Show-EventLog,Show-WindowsDeveloperLicenseRegistration,Sort-Object,Split-Path,Split-WindowsImage,Start-BitsTransfer,Start-DscConfiguration,Start-DtiagnosticResourceManager,Start-Job,Start-Process,Start-Service,Start-Sleep,Start-Transaction,Start-Transcript,Stop-Computer,Stop-DtiagnosticResourceManager,Stop-Job,Stop-Process,Stop-Service,Stop-Transcript,Suspend-BitsTransfer,Suspend-Job,Suspend-Service,Switch-Certificate,Tee-Object,Test-Certificate,Test-ComputerSecureChannel,Test-Connection,Test-DscConfiguration,Test-FileCatalog,Test-KdsRootKey,Test-ModuleManifest,Test-Path,Test-PSSessionConfigurationFile,Test-WSMan,Trace-Command,Unblock-File,Unblock-Tpm,Undo-DtiagnosticTransaction,Undo-Transaction,Uninstall-Package,Uninstall-ProvisioningPackage,Uninstall-TrustedProvisioningCertificate,Unprotect-CmsMessage,Unregister-Event,Unregister-PackageSource,Unregister-PSSessionConfiguration,Unregister-ScheduledJob,Unregister-WindowsDeveloperLicense,Update-FormatData,Update-Help,Update-List,Update-TypeData,Update-WIMBootEntry,Use-Transaction,Use-WindowsUnattend,Wait-Debugger,Wait-Event,Wait-Job,Wait-Process,Where-Object,Write-Debug,Write-Error,Write-EventLog,Write-Host,Write-Information,Write-Output,Write-Progress,Write-Verbose,Write-Warning
	},
	morekeywords={
		Add-BitLockerKeyProtector,Add-DnsClientNrptRule,Add-DtcClusterTMMapping,Add-EtwTraceProvider,Add-InitiatorIdToMaskingSet,Add-MpPreference,Add-NetEventNetworkAdapter,Add-NetEventPacketCaptureProvider,Add-NetEventProvider,Add-NetEventVFPProvider,Add-NetEventVmNetworkAdapter,Add-NetEventVmSwitch,Add-NetEventVmSwitchProvider,Add-NetEventWFPCaptureProvider,Add-NetIPHttpsCertBinding,Add-NetLbfoTeamMember,Add-NetLbfoTeamNic,Add-NetNatExternalAddress,Add-NetNatStaticMapping,Add-NetSwitchTeamMember,Add-Odbsn,Add-PartitionAccessPath,Add-PhysicalDisk,Add-Printer,Add-PrinterDriver,Add-PrinterPort,Add-StorageFaultDomain,Add-TargetPortToMaskingSet,Add-VirtualDiskToMaskingSet,Add-VpnConnection,Add-VpnConnectionRoute,Add-VpnConnectionTriggerApplication,Add-VpnConnectionTriggerDnsConfiguration,Add-VpnConnectionTriggerTrustedNetwork,AfterAll,AfterEach,Assert-MockCalled,Assert-VerifiableMocks,Backup-BitLockerKeyProtector,BackupToAAD-BitLockerKeyProtector,BeforeAll,BeforeEach,Block-FileShareAccess,Block-SmbShareAccess,Clear-BitLockerAutoUnlock,Clear-Disk,Clear-DnsClientCache,Clear-FileStorageTier,Clear-Host,Clear-PcsvDeviceLog,Clear-StorageDiagnosticInfo,Close-SmbOpenFile,Close-SmbSession,Compress-Archive,Configuration,Connect-IscsiTarget,Connect-VirtualDisk,Context,convert,ConvertFrom-SddlString,Copy-NetFirewallRule,Copy-NetIPsecMainModeCryptoSet,Copy-NetIPsecMainModeRule,Copy-NetIPsecPhase1AuthSet,Copy-NetIPsecPhase2AuthSet,Copy-NetIPsecQuickModeCryptoSet,Copy-NetIPsecRule,Debug-FileShare,Debug-MMAppPrelaunch,Debug-StorageSubSystem,Debug-Volume,Describe,Disable-BitLocker,Disable-BitLockerAutoUnlock,Disable-DAManualEntryPointSelection,Disable-Dsebug,Disable-MMAgent,Disable-NetAdapter,Disable-NetAdapterBinding,Disable-NetAdapterChecksumOffload,Disable-NetAdapterEncapsulatedPacketTaskOffload,Disable-NetAdapterIPsecOffload,Disable-NetAdapterLso,Disable-NetAdapterPacketDirect,Disable-NetAdapterPowerManagement,Disable-NetAdapterQos,Disable-NetAdapterRdma,Disable-NetAdapterRsc,Disable-NetAdapterRss,Disable-NetAdapterSriov,Disable-NetAdapterVmq,Disable-NetDnsTransitionConfiguration,Disable-NetFirewallRule,Disable-NetIPHttpsProfile,Disable-NetIPsecMainModeRule,Disable-NetIPsecRule,Disable-NetNatTransitionConfiguration,Disable-NetworkSwitchEthernetPort,Disable-NetworkSwitchFeature,Disable-NetworkSwitchVlan,Disable-OdbcPerfCounter,Disable-PhysicalDiskIdentification,Disable-PnpDevice,Disable-PSTrace,Disable-PSWSManCombinedTrace,Disable-ScheduledTask,Disable-SmbDelegation,Disable-StorageEnclosureIdentification,Disable-StorageEnclosurePower,Disable-StorageHighAvailability,Disable-StorageMaintenanceMode,Disable-WdacBidTrace,Disable-WSManTrace,Disconnect-IscsiTarget,Disconnect-VirtualDisk,Dismount-DiskImage,Enable-BitLocker,Enable-BitLockerAutoUnlock,Enable-DAManualEntryPointSelection,Enable-Dsebug,Enable-MMAgent,Enable-NetAdapter,Enable-NetAdapterBinding,Enable-NetAdapterChecksumOffload,Enable-NetAdapterEncapsulatedPacketTaskOffload,Enable-NetAdapterIPsecOffload,Enable-NetAdapterLso,Enable-NetAdapterPacketDirect,Enable-NetAdapterPowerManagement,Enable-NetAdapterQos,Enable-NetAdapterRdma,Enable-NetAdapterRsc,Enable-NetAdapterRss,Enable-NetAdapterSriov,Enable-NetAdapterVmq,Enable-NetDnsTransitionConfiguration,Enable-NetFirewallRule,Enable-NetIPHttpsProfile,Enable-NetIPsecMainModeRule,Enable-NetIPsecRule,Enable-NetNatTransitionConfiguration,Enable-NetworkSwitchEthernetPort,Enable-NetworkSwitchFeature,Enable-NetworkSwitchVlan,Enable-OdbcPerfCounter,Enable-PhysicalDiskIdentification,Enable-PnpDevice,Enable-PSTrace,Enable-PSWSManCombinedTrace,Enable-ScheduledTask,Enable-SmbDelegation,Enable-StorageEnclosureIdentification,Enable-StorageEnclosurePower,Enable-StorageHighAvailability,Enable-StorageMaintenanceMode,Enable-WdacBidTrace,Enable-WSManTrace,Expand-Archive,Export-ODataEndpointProxy,Export-ScheduledTask,Find-Command,Find-DscResource,Find-Module,Find-NetIPsecRule,Find-NetRoute,Find-RoleCapability,Find-Script,Flush-EtwTraceSession,Format-Hex,Format-Volume,Get-AppBackgroundTask,Get-AppxLastError,Get-AppxLog,Get-AutologgerConfig,Get-BitLockerVolume,Get-ClusteredScheduledTask,Get-DAClientExperienceConfiguration,Get-DAConnectionStatus,Get-DAEntryPointTableItem,Get-DedupProperties,Get-Disk,Get-DiskImage,Get-DiskStorageNodeView,Get-DnsClient,Get-DnsClientCache,Get-DnsClientGlobalSetting,Get-DnsClientNrptGlobal,Get-DnsClientNrptPolicy,Get-DnsClientNrptRule,Get-DnsClientServerAddress,Get-DscConfiguration,Get-DscConfigurationStatus,Get-DscLocalConfigurationManager,Get-DscResource,Get-Dtc,Get-DtcAdvancedHostSetting,Get-DtcAdvancedSetting,Get-DtcClusterDefault,Get-DtcClusterTMMapping,Get-Dtefault,Get-DtcLog,Get-DtcNetworkSetting,Get-DtcTransaction,Get-DtcTransactionsStatistics,Get-DtcTransactionsTraceSession,Get-DtcTransactionsTraceSetting,Get-EtwTraceProvider,Get-EtwTraceSession,Get-FileHash,Get-FileIntegrity,Get-FileShare,Get-FileShareAccessControlEntry,Get-FileStorageTier,Get-InitiatorId,Get-InitiatorPort,Get-InstalledModule,Get-InstalledScript,Get-IscsiConnection,Get-IscsiSession,Get-IscsiTarget,Get-IscsiTargetPortal,Get-IseSnippet,Get-LogProperties,Get-MaskingSet,Get-MMAgent,Get-MockDynamicParameters,Get-MpComputerStatus,Get-MpPreference,Get-MpThreat,Get-MpThreatCatalog,Get-MpThreatDetection,Get-NCSIPolicyConfiguration,Get-Net6to4Configuration,Get-NetAdapter,Get-NetAdapterAdvancedProperty,Get-NetAdapterBinding,Get-NetAdapterChecksumOffload,Get-NetAdapterEncapsulatedPacketTaskOffload,Get-NetAdapterHardwareInfo,Get-NetAdapterIPsecOffload,Get-NetAdapterLso,Get-NetAdapterPacketDirect,Get-NetAdapterPowerManagement,Get-NetAdapterQos,Get-NetAdapterRdma,Get-NetAdapterRsc,Get-NetAdapterRss,Get-NetAdapterSriov,Get-NetAdapterSriovVf,Get-NetAdapterStatistics,Get-NetAdapterVmq,Get-NetAdapterVMQQueue,Get-NetAdapterVPort,Get-NetCompartment,Get-NetConnectionProfile,Get-NetDnsTransitionConfiguration,Get-NetDnsTransitionMonitoring,Get-NetEventNetworkAdapter,Get-NetEventPacketCaptureProvider,Get-NetEventProvider,Get-NetEventSession,Get-NetEventVFPProvider,Get-NetEventVmNetworkAdapter,Get-NetEventVmSwitch,Get-NetEventVmSwitchProvider,Get-NetEventWFPCaptureProvider,Get-NetFirewallAddressFilter,Get-NetFirewallApplicationFilter,Get-NetFirewallInterfaceFilter,Get-NetFirewallInterfaceTypeFilter,Get-NetFirewallPortFilter,Get-NetFirewallProfile,Get-NetFirewallRule,Get-NetFirewallSecurityFilter,Get-NetFirewallServiceFilter,Get-NetFirewallSetting,Get-NetIPAddress,Get-NetIPConfiguration,Get-NetIPHttpsConfiguration,Get-NetIPHttpsState,Get-NetIPInterface,Get-NetIPseospSetting,Get-NetIPsecMainModeCryptoSet,Get-NetIPsecMainModeRule,Get-NetIPsecMainModeSA,Get-NetIPsecPhase1AuthSet,Get-NetIPsecPhase2AuthSet,Get-NetIPsecQuickModeCryptoSet,Get-NetIPsecQuickModeSA,Get-NetIPsecRule,Get-NetIPv4Protocol,Get-NetIPv6Protocol,Get-NetIsatapConfiguration,Get-NetLbfoTeam,Get-NetLbfoTeamMember,Get-NetLbfoTeamNic,Get-NetNat,Get-NetNatExternalAddress,Get-NetNatGlobal,Get-NetNatSession,Get-NetNatStaticMapping,Get-NetNatTransitionConfiguration,Get-NetNatTransitionMonitoring,Get-NetNeighbor,Get-NetOffloadGlobalSetting,Get-NetPrefixPolicy,Get-NetQosPolicy,Get-NetRoute,Get-NetSwitchTeam,Get-NetSwitchTeamMember,Get-NetTCPConnection,Get-NetTCPSetting,Get-NetTeredoConfiguration,Get-NetTeredoState,Get-NetTransportFilter,Get-NetUDPEndpoint,Get-NetUDPSetting,Get-NetworkSwitchEthernetPort,Get-NetworkSwitchFeature,Get-NetworkSwitchGlobalData,Get-NetworkSwitchVlan,Get-Odbriver,Get-Odbsn,Get-OdbcPerfCounter,Get-OffloadDataTransferSetting,Get-OperationValidation,Get-Partition,Get-PartitionSupportedSize,Get-PcsvDevice,Get-PcsvDeviceLog,Get-PhysicalDisk,Get-PhysicalDiskStorageNodeView,Get-PhysicalExtent,Get-PhysicalExtentAssociation,Get-PnpDevice,Get-PnpDeviceProperty,Get-PrintConfiguration,Get-Printer,Get-PrinterDriver,Get-PrinterPort,Get-PrinterProperty,Get-PrintJob,Get-PSRepository,Get-ResiliencySetting,Get-ScheduledTask,Get-ScheduledTaskInfo,Get-SmbBandWidthLimit,Get-SmbClientConfiguration,Get-SmbClientNetworkInterface,Get-SmbConnection,Get-SmbDelegation,Get-SmbGlobalMapping,Get-SmbMapping,Get-SmbMultichannelConnection,Get-SmbMultichannelConstraint,Get-SmbOpenFile,Get-SmbServerConfiguration,Get-SmbServerNetworkInterface,Get-SmbSession,Get-SmbShare,Get-SmbShareAccess,Get-SmbWitnessClient,Get-StartApps,Get-StorageAdvancedProperty,Get-StorageDiagnosticInfo,Get-StorageEnclosure,Get-StorageEnclosureStorageNodeView,Get-StorageEnclosureVendorData,Get-StorageExtendedStatus,Get-StorageFaultDomain,Get-StorageFileServer,Get-StorageFirmwareInformation,Get-StorageHealthAction,Get-StorageHealthReport,Get-StorageHealthSetting,Get-StorageJob,Get-StorageNode,Get-StoragePool,Get-StorageProvider,Get-StorageReliabilityCounter,Get-StorageSetting,Get-StorageSubSystem,Get-StorageTier,Get-StorageTierSupportedSize,Get-SupportedClusterSizes,Get-SupportedFileSystems,Get-TargetPort,Get-TargetPortal,Get-TestDriveItem,Get-Verb,Get-VirtualDisk,Get-VirtualDiskSupportedSize,Get-Volume,Get-VolumeCorruptionCount,Get-VolumeScrubPolicy,Get-VpnConnection,Get-VpnConnectionTrigger,Get-WdacBidTrace,Get-WindowsUpdateLog,Get-WUAVersion,Get-WUIsPendingReboot,Get-WULastInstallationDate,Get-WULastScanSuccessDate,Grant-FileShareAccess,Grant-SmbShareAccess,help,Hide-VirtualDisk,Import-IseSnippet,Import-PowerShellDataFile,ImportSystemModules,In,Initialize-Disk,InModuleScope,Install-Dtc,Install-Module,Install-Script,Install-WUUpdates,Invoke-AsWorkflow,Invoke-Mock,Invoke-OperationValidation,Invoke-Pester,It,Lock-BitLocker,mkdir,Mock,more,Mount-DiskImage,Move-SmbWitnessClient,New-AutologgerConfig,New-DAEntryPointTableItem,New-DscChecksum,New-EapConfiguration,New-EtwTraceSession,New-FileShare,New-Fixture,New-Guid,New-IscsiTargetPortal,New-IseSnippet,New-MaskingSet,New-NetAdapterAdvancedProperty,New-NetEventSession,New-NetFirewallRule,New-NetIPAddress,New-NetIPHttpsConfiguration,New-NetIPseospSetting,New-NetIPsecMainModeCryptoSet,New-NetIPsecMainModeRule,New-NetIPsecPhase1AuthSet,New-NetIPsecPhase2AuthSet,New-NetIPsecQuickModeCryptoSet,New-NetIPsecRule,New-NetLbfoTeam,New-NetNat,New-NetNatTransitionConfiguration,New-NetNeighbor,New-NetQosPolicy,New-NetRoute,New-NetSwitchTeam,New-NetTransportFilter,New-NetworkSwitchVlan,New-Partition,New-PesterOption,New-PSWorkflowSession,New-ScheduledTask,New-ScheduledTaskAction,New-ScheduledTaskPrincipal,New-ScheduledTaskSettingsSet,New-ScheduledTaskTrigger,New-ScriptFileInfo,New-SmbGlobalMapping,New-SmbMapping,New-SmbMultichannelConstraint,New-SmbShare,New-StorageFileServer,New-StoragePool,New-StorageSubsystemVirtualDisk,New-StorageTier,New-TemporaryFile,New-VirtualDisk,New-VirtualDiskClone,New-VirtualDiskSnapshot,New-Volume,New-VpnServerAddress,Open-NetGPO,Optimize-StoragePool,Optimize-Volume,oss,Pause,prompt,PSConsoleHostReadline,Publish-Module,Publish-Script,Read-PrinterNfcTag,Register-ClusteredScheduledTask,Register-DnsClient,Register-IscsiSession,Register-PSRepository,Register-ScheduledTask,Register-StorageSubsystem,Remove-AutologgerConfig,Remove-BitLockerKeyProtector,Remove-DAEntryPointTableItem,Remove-DnsClientNrptRule,Remove-DscConfigurationDocument,Remove-DtcClusterTMMapping,Remove-EtwTraceProvider,Remove-FileShare,Remove-InitiatorId,Remove-InitiatorIdFromMaskingSet,Remove-IscsiTargetPortal,Remove-MaskingSet,Remove-MpPreference,Remove-MpThreat,Remove-NetAdapterAdvancedProperty,Remove-NetEventNetworkAdapter,Remove-NetEventPacketCaptureProvider,Remove-NetEventProvider,Remove-NetEventSession,Remove-NetEventVFPProvider,Remove-NetEventVmNetworkAdapter,Remove-NetEventVmSwitch,Remove-NetEventVmSwitchProvider,Remove-NetEventWFPCaptureProvider,Remove-NetFirewallRule,Remove-NetIPAddress,Remove-NetIPHttpsCertBinding,Remove-NetIPHttpsConfiguration,Remove-NetIPseospSetting,Remove-NetIPsecMainModeCryptoSet,Remove-NetIPsecMainModeRule,Remove-NetIPsecMainModeSA,Remove-NetIPsecPhase1AuthSet,Remove-NetIPsecPhase2AuthSet,Remove-NetIPsecQuickModeCryptoSet,Remove-NetIPsecQuickModeSA,Remove-NetIPsecRule,Remove-NetLbfoTeam,Remove-NetLbfoTeamMember,Remove-NetLbfoTeamNic,Remove-NetNat,Remove-NetNatExternalAddress,Remove-NetNatStaticMapping,Remove-NetNatTransitionConfiguration,Remove-NetNeighbor,Remove-NetQosPolicy,Remove-NetRoute,Remove-NetSwitchTeam,Remove-NetSwitchTeamMember,Remove-NetTransportFilter,Remove-NetworkSwitchEthernetPortIPAddress,Remove-NetworkSwitchVlan,Remove-Odbsn,Remove-Partition,Remove-PartitionAccessPath,Remove-PhysicalDisk,Remove-Printer,Remove-PrinterDriver,Remove-PrinterPort,Remove-PrintJob,Remove-SmbBandwidthLimit,Remove-SmbGlobalMapping,Remove-SmbMapping,Remove-SmbMultichannelConstraint,Remove-SmbShare,Remove-StorageFaultDomain,Remove-StorageFileServer,Remove-StorageHealthIntent,Remove-StorageHealthSetting,Remove-StoragePool,Remove-StorageTier,Remove-TargetPortFromMaskingSet,Remove-VirtualDisk,Remove-VirtualDiskFromMaskingSet,Remove-VpnConnection,Remove-VpnConnectionRoute,Remove-VpnConnectionTriggerApplication,Remove-VpnConnectionTriggerDnsConfiguration,Remove-VpnConnectionTriggerTrustedNetwork,Rename-DAEntryPointTableItem,Rename-MaskingSet,Rename-NetAdapter,Rename-NetFirewallRule,Rename-NetIPHttpsConfiguration,Rename-NetIPsecMainModeCryptoSet,Rename-NetIPsecMainModeRule,Rename-NetIPsecPhase1AuthSet,Rename-NetIPsecPhase2AuthSet,Rename-NetIPsecQuickModeCryptoSet,Rename-NetIPsecRule,Rename-NetLbfoTeam,Rename-NetSwitchTeam,Rename-Printer,Repair-FileIntegrity,Repair-VirtualDisk,Repair-Volume,Reset-DAClientExperienceConfiguration,Reset-DAEntryPointTableItem,Reset-DtcLog,Reset-NCSIPolicyConfiguration,Reset-Net6to4Configuration,Reset-NetAdapterAdvancedProperty,Reset-NetDnsTransitionConfiguration,Reset-NetIPHttpsConfiguration,Reset-NetIsatapConfiguration,Reset-NetTeredoConfiguration,Reset-PhysicalDisk,Reset-StorageReliabilityCounter,Resize-Partition,Resize-StorageTier,Resize-VirtualDisk,Restart-NetAdapter,Restart-PcsvDevice,Restart-PrintJob,Restore-DscConfiguration,Restore-NetworkSwitchConfiguration,Resume-BitLocker,Resume-PrintJob,Revoke-FileShareAccess,Revoke-SmbShareAccess,SafeGetCommand,Save-EtwTraceSession,Save-Module,Save-NetGPO,Save-NetworkSwitchConfiguration,Save-Script,Send-EtwTraceSession,Set-AutologgerConfig,Set-ClusteredScheduledTask,Set-DAClientExperienceConfiguration,Set-DAEntryPointTableItem,Set-Disk,Set-DnsClient,Set-DnsClientGlobalSetting,Set-DnsClientNrptGlobal,Set-DnsClientNrptRule,Set-DnsClientServerAddress,Set-DtcAdvancedHostSetting,Set-DtcAdvancedSetting,Set-DtcClusterDefault,Set-DtcClusterTMMapping,Set-Dtefault,Set-DtcLog,Set-DtcNetworkSetting,Set-DtcTransaction,Set-DtcTransactionsTraceSession,Set-DtcTransactionsTraceSetting,Set-DynamicParameterVariables,Set-EtwTraceProvider,Set-FileIntegrity,Set-FileShare,Set-FileStorageTier,Set-InitiatorPort,Set-IscsiChapSecret,Set-LogProperties,Set-MMAgent,Set-MpPreference,Set-NCSIPolicyConfiguration,Set-Net6to4Configuration,Set-NetAdapter,Set-NetAdapterAdvancedProperty,Set-NetAdapterBinding,Set-NetAdapterChecksumOffload,Set-NetAdapterEncapsulatedPacketTaskOffload,Set-NetAdapterIPsecOffload,Set-NetAdapterLso,Set-NetAdapterPacketDirect,Set-NetAdapterPowerManagement,Set-NetAdapterQos,Set-NetAdapterRdma,Set-NetAdapterRsc,Set-NetAdapterRss,Set-NetAdapterSriov,Set-NetAdapterVmq,Set-NetConnectionProfile,Set-NetDnsTransitionConfiguration,Set-NetEventPacketCaptureProvider,Set-NetEventProvider,Set-NetEventSession,Set-NetEventVFPProvider,Set-NetEventVmSwitchProvider,Set-NetEventWFPCaptureProvider,Set-NetFirewallAddressFilter,Set-NetFirewallApplicationFilter,Set-NetFirewallInterfaceFilter,Set-NetFirewallInterfaceTypeFilter,Set-NetFirewallPortFilter,Set-NetFirewallProfile,Set-NetFirewallRule,Set-NetFirewallSecurityFilter,Set-NetFirewallServiceFilter,Set-NetFirewallSetting,Set-NetIPAddress,Set-NetIPHttpsConfiguration,Set-NetIPInterface,Set-NetIPseospSetting,Set-NetIPsecMainModeCryptoSet,Set-NetIPsecMainModeRule,Set-NetIPsecPhase1AuthSet,Set-NetIPsecPhase2AuthSet,Set-NetIPsecQuickModeCryptoSet,Set-NetIPsecRule,Set-NetIPv4Protocol,Set-NetIPv6Protocol,Set-NetIsatapConfiguration,Set-NetLbfoTeam,Set-NetLbfoTeamMember,Set-NetLbfoTeamNic,Set-NetNat,Set-NetNatGlobal,Set-NetNatTransitionConfiguration,Set-NetNeighbor,Set-NetOffloadGlobalSetting,Set-NetQosPolicy,Set-NetRoute,Set-NetTCPSetting,Set-NetTeredoConfiguration,Set-NetUDPSetting,Set-NetworkSwitchEthernetPortIPAddress,Set-NetworkSwitchPortMode,Set-NetworkSwitchPortProperty,Set-NetworkSwitchVlanProperty,Set-Odbriver,Set-Odbsn,Set-Partition,Set-PcsvDeviceBootConfiguration,Set-PcsvDeviceNetworkConfiguration,Set-PcsvDeviceUserPassword,Set-PhysicalDisk,Set-PrintConfiguration,Set-Printer,Set-PrinterProperty,Set-PSRepository,Set-ResiliencySetting,Set-ScheduledTask,Set-SmbBandwidthLimit,Set-SmbClientConfiguration,Set-SmbPathAcl,Set-SmbServerConfiguration,Set-SmbShare,Set-StorageFileServer,Set-StorageHealthSetting,Set-StoragePool,Set-StorageProvider,Set-StorageSetting,Set-StorageSubSystem,Set-StorageTier,Set-TestInconclusive,Setup,Set-VirtualDisk,Set-Volume,Set-VolumeScrubPolicy,Set-VpnConnection,Set-VpnConnectionIPsecConfiguration,Set-VpnConnectionProxy,Set-VpnConnectionTriggerDnsConfiguration,Set-VpnConnectionTriggerTrustedNetwork,Should,Show-NetFirewallRule,Show-NetIPsecRule,Show-VirtualDisk,Start-AppBackgroundTask,Start-AutologgerConfig,Start-Dtc,Start-DtcTransactionsTraceSession,Start-EtwTraceSession,Start-MpScan,Start-MpWDOScan,Start-NetEventSession,Start-PcsvDevice,Start-ScheduledTask,Start-StorageDiagnosticLog,Start-Trace,Start-WUScan,Stop-DscConfiguration,Stop-Dtc,Stop-DtcTransactionsTraceSession,Stop-EtwTraceSession,Stop-NetEventSession,Stop-PcsvDevice,Stop-ScheduledTask,Stop-StorageDiagnosticLog,Stop-StorageJob,Stop-Trace,Suspend-BitLocker,Suspend-PrintJob,Sync-NetIPsecRule,TabExpansion2,Test-Dtc,Test-NetConnection,Test-ScriptFileInfo,Unblock-FileShareAccess,Unblock-SmbShareAccess,Uninstall-Dtc,Uninstall-Module,Uninstall-Script,Unlock-BitLocker,Unregister-AppBackgroundTask,Unregister-ClusteredScheduledTask,Unregister-IscsiSession,Unregister-PSRepository,Unregister-ScheduledTask,Unregister-StorageSubsystem,Update-Disk,Update-DscConfiguration,Update-EtwTraceSession,Update-HostStorageCache,Update-IscsiTarget,Update-IscsiTargetPortal,Update-Module,Update-ModuleManifest,Update-MpSignature,Update-NetIPsecRule,Update-Script,Update-ScriptFileInfo,Update-SmbMultichannelConnection,Update-StorageFirmware,Update-StoragePool,Update-StorageProviderCache,Write-DtcTransactionsTraceSession,Write-PrinterNfcTag,Write-VolumeCache
	},
	morekeywords={Do,Else,For,ForEach,Function,If,In,Until,While},
	alsodigit={-},
	sensitive=false,
	morecomment=[l]{\#},
	morecomment=[n]{<\#}{\#>},
	morestring=[b]{"},
	morestring=[b]{'},
	morestring=[s]{@'}{'@},
	morestring=[s]{@"}{"@}
}
\definecolor{dkgreen}{rgb}{0,0.6,0}
\definecolor{mauve}{rgb}{0.88, 0.69, 1.0}
\lstdefinelanguage{batch}{
    morekeywords={not,exist,mkdir,echo,pause},
    morekeywords={Do,Else,For,ForEach,Function,If,In,Until,While},
    alsodigit={-},
    sensitive=false,
    morecomment=[l]{\#},
    morecomment=[n]{<\#}{\#>},
    morestring=[b]{"},
    morestring=[b]{'},
    morestring=[s]{@'}{'@},
    morestring=[s]{@"}{"@},
    numbers=left,
    numberstyle=\tiny,
    basicstyle={\small\ttfamily},
    stringstyle=\color{dkgreen},
    keywordstyle=\color{blue},
    frame=tb,
}

\makeatletter

\newcommand*{\overrightharpoonup}{\mathpalette{\overarrow@\rightharpoonupfill@}}
\newcommand*{\rightharpoonupfill@}{\arrowfill@\relbar\relbar\rightharpoonup}
\makeatother

\makeatletter
\newcommand*{\overleftharpoonup}{\mathpalette{\overarrow@\leftharpoonupfill@}}
\newcommand*{\leftharpoonupfill@}{\arrowfill@\relbar\relbar\leftharpoonup}
\makeatother

\definecolor{gray90}{gray}{0.9}

\makeatletter
\newcommand*\bigcdot{\mathpalette\bigcdot@{1}}
\newcommand*\bigcdot@[2]{\mathbin{\vcenter{\hbox{\scalebox{#2}{$\m@th#1\bullet$}}}}}
\makeatother

\crefname{section}{Section}{Sections}
\crefname{equation}{Equation}{Equations}
\crefname{figure}{Figure}{Figures}
\crefname{table}{Table}{Tables}
\crefname{listing}{Listing}{Listings}

\setlist[itemize]{leftmargin=10pt,itemindent=0pt,topsep=2pt,partopsep=2.5pt,parsep=1pt,itemsep=1.5pt,listparindent=\parindent{}}
\setlist[enumerate]{leftmargin=16pt,itemindent=0pt,topsep=2pt,partopsep=2.5pt,parsep=1pt,itemsep=1.5pt,listparindent=\parindent{}}

\title{Evolution of NVENC Efficiency: A Longitudinal Analysis of HQ and UHQ Tuning Efficiency, Latency and Energy Trade-offs}
\name{Kasidis Arunruangsirilert, Jiro Katto}
\address{School of Fundamental Science and Engineering, Waseda University, Tokyo, Japan}

\begin{document}
\bstctlcite{IEEEexample:BSTcontrol}
%
\maketitle

\setstretch{0.941}
\begin{abstract}

The rapid expansion of uplink-intensive applications necessitates video coding solutions that balance high Rate-Distortion (RD) efficiency with ultra-low latency. This paper presents a longitudinal performance analysis of NVIDIA hardware encoding (NVENC), spanning from Pascal to the emerging Blackwell generation. We specifically evaluate the operational viability of the new ``Ultra High Quality" (UHQ) tuning mode against standard low-latency configurations. Our results demonstrate that while the Blackwell architecture breaks historical efficiency plateaus, achieving a 5.94\% BD-Rate gain in standard modes and up to 22.79\% in UHQ modes, these gains incur severe system-level penalties. We reveal that UHQ operates as a hybrid pipeline, offloading complexity to CUDA cores and enforcing aggressive temporal structures (up to 7 B-frames) that increase end-to-end latency by over 400\% and GPU board power consumption by up to 40\%. Consequently, while UHQ successfully bridges the quality gap with software encoders, its prohibitive serialization delay renders it unsuitable for interactive real-time communications, positioning it instead as a specialized solution for Video-on-Demand (VoD) transcoding.

\end{abstract}
%
\vspace{-3mm}
\section{Introduction}
\label{sec:intro}
\vspace{-2mm}

\blfootnote{This paper is supported by the Ministry of Internal Affairs and Communications (MIC) Project for Efficient Frequency Utilization Toward Wireless IP Multicasting and the Japan Science and Technology Agency (JST) CRONOS Grant Number JPMJCS25N2.}

The proliferation of immersive media formats, such as Volumetric Video-based Point Cloud Compression (V-PCC) \cite{10.1145/3682062,10.1145/3690641,8966190} and high-fidelity real-time streaming \cite{9043021, 9825297}, places unprecedented strain on network uplink capacities. While 5G standards define multi-gigabit throughput, practical deployments are fundamentally constrained by Time Division Duplex (TDD) configurations. Most commercial 5G networks utilize a 7:2 downlink-to-uplink slot ratio to prioritize consumption traffic, severely limiting the available bandwidth for uplink transmission \cite{10570635}. Although mmWave (Frequency Range 2 or FR2) offers high uplink throughput, due to its wider bandwidth, its propagation characteristics restrict coverage to dense urban pockets, leaving the majority of users reliant on Sub-6 (FR1) networks with constrained uplink budgets \cite{11161113, 11366446}. This structural asymmetry necessitates video coding solutions that maximize Rate-Distortion (RD) efficiency without incurring the latency and power consumption penalty of the software-based encoding. 

To cope with these challenges, Hardware-accelerated encoders, such as NVIDIA NVENC, Intel QuickSync, and AMD VCE, have been integrated into the die of modern Graphics Processing Units (GPUs) to deal with video coding-related tasks and have become the standard for addressing these latency constraints. Historically, these fixed-function Silicon Intellectual Property (SIP) blocks prioritized throughput over coding efficiency. However, our previous work found that modern SIP achieves compression efficiency that is on par or slightly exceeds that of software encoders in real-time encoding scenarios \cite{10637525}. Additionally, recent architectural trends indicate a shift toward hybrid encoding pipelines. 

Recent hardware iterations, such as the NVIDIA Ada Lovelace and Blackwell architectures, introduce ``Ultra High Quality" (UHQ) tuning modes that purportedly leverage general-purpose Streaming Multiprocessor (SM) or CUDA cores to augment the fixed-function encoder. While vendor documentation suggests these hybrid modes can achieve BD-Rate improvements of up to 15\% compared to standard ``High-Quality" configurations \cite{nvidia_bw_2024, nvidia_2024b}, the system-level implications of this design shift remain under-explored in the literature. Furthermore, Blackwell introduces native 10-bit internal processing for HEVC, allowing 8-bit content to be encoded with higher precision to achieve a claimed 3\% compression efficiency gain. While earlier architectures (Ada Lovelace) supported similar features via CUDA-based pre-processing, the shift to native hardware implementation in Blackwell represents a distinct micro-architectural divergence designed to improve efficiency without the associated computational overhead \cite{vishesh_lokras_2025, nvidia_2026}.

However, the specific mechanisms underlying the UHQ mode remain undocumented. While NVIDIA documentation suggested a hybrid pipeline, it does not explicitly define the specific algorithmic modifications employed to achieve the reported gains. The term ``Ultra High Quality" serves as an abstraction layer, potentially masking aggressive design choices, such as extended Group of Pictures (GOP) structure or deep B-frame hierarchies, that are critical to playback compatibility and decoding performance. Moreover, activating UHQ mode explicitly locks out manual control over key encoding parameters, such as the number of reference frames and B-frames. This forced override prevents the users from tuning the encoder for specific latency constraints, effectively mandating a ``black box" configuration, to maintain the claimed compression efficiency number. Without granular visibility into these locked parameters, it is impossible to determine whether the efficiency gains stem from genuine architectural improvements or simply from imposing computationally expensive temporal structures that may violate strict motion-to-photon latency thresholds.

\begin{table}[!tbp]
\vspace{-2.5mm}
\setstretch{0.75}
\caption{Hardware and Software Configuration}
\centering
\label{tab:hardware}
\resizebox{8.5cm}{!}{\begin{tabular}{@{}ll@{}}
\toprule
\multicolumn{2}{c}{Encoding Systems}\\
\midrule
Hardware                 & Description  \\\midrule
Pascal GPU& NVIDIA GeForce GTX 1070 (GP104, 
314 mm²)\\
Ampere GPU& NVIDIA GeForce RTX 3060 12 GB (GA104, 392 mm²)\\
Ada Lovelace GPU& NVIDIA GeForce RTX 4070 Ti SUPER (AD103, 379 mm²)\\
Blackwell GPU& NVIDIA GeForce RTX 5070 Ti (GB203, 378 mm²)\\\midrule
Software & Version \\\midrule
OS & Microsoft Windows 10/11/Server 2025 Datacenter\\
ffmpeg & N-122268-g0dfaed77a6\\
OBS Studio & 32.0.4\\
NVIDIA GPU Driver & GeForce Game Ready Driver 581.80 (581.80)\\\midrule
\multicolumn{2}{c}{VMAF/PSNR Calculation System}\\
\midrule
Hardware                 & Description  \\\midrule
CPU & Intel(R) Core(TM) Ultra 9 285K \\
RAM & Dual-Channel DDR5 128 GB (4×32 GB) @ 4400 MT/s \\ 
GPU & NVIDIA RTX PRO 5000 48GB Blackwell\\\midrule
Software & Version \\\midrule
OS & Ubuntu 24.04.3 LTS\\
ffmpeg & N-122271-g0629780cf6\\
libvmaf & v3.0.0 (b9ac69e6)\\
VMAF Model & vmaf\_4k\_v0.6.1neg \\
NVIDIA GPU Driver & 580.126.09 \\
NVIDIA CUDA Compiler & cuda\_13.1.r13.1\/compiler.36836380\_0 \\
\midrule
\multicolumn{2}{c}{WebRTC Server}\\
\midrule
Hardware                 & Description  \\\midrule
System Board & Dell PowerEdge R7515\\
CPU & AMD EPYC 7C13 64-Core Processor\\
RAM & Octal-Channel DDR4 512 GB @ 2666 MT/s \\\midrule
Software & Version \\\midrule
OS & Ubuntu 24.04.1 LTS\\
WebRTC Server & SRS 6.0.101\\
\bottomrule
\end{tabular}}
\vspace{-7mm}
\end{table}

In this paper, we present a longitudinal performance analysis of NVIDIA NVENC, spanning from the Pascal architecture (10-series) to the emerging Blackwell architecture (50-series). We rigorously evaluate the trade-offs between RD Performance, Encoding Throughput, End-to-End (E2E) Latency, and Power Consumption. Specifically, we investigated:
\begin{itemize}
    \item \textbf{Efficiency vs. Complexity:} We measure the BD-Rate evolution of HEVC and AV1 across generations, isolating the impact of UHQ tuning and 10-bit internal processing against standard baselines.
    \item \textbf{System-Level Costs:} We quantify the hidden costs of these tuning modes, specifically measuring the latency penalty and raw power consumption associated with UHQ usage.
    \item \textbf{Deconstructing UHQ:} We reverse-engineer the operational behavior of the UHQ mode, analyzing the resulting bitstream characteristics (B-frame depth, active reference utilization) to understand the specific algorithmic trade-offs NVIDIA has implemented.
\end{itemize}
\vspace{-3.5mm}
\section{Experiment Setup}
\vspace{-1mm}
\subsection{Hardware Selection}

To strictly focus on the generational evolution of NVENC, we selected GPUs with the Pascal, Ampere, Ada Lovelace, and Blackwell architectures, specifically targeting the 70-class performance tier to maintain comparable die sizes and thermal characteristics (see \cref{tab:hardware}). For the Ampere generation, due to procurement constraints regarding the RTX 3070, we utilized a specific variant of the RTX 3060 12 GB manufactured using the GA104 die, the same silicon found in the RTX 3070, thereby preserving the validity of the power and thermal comparisons. Moreover, we excluded the Turing architectures (20-series), as their HEVC encoder is functionally identical to the Ampere generation; the primary differentiation lies in Ampere's addition of AV1 decoding support, which is outside the scope of this encoding-focused benchmark \cite{muthana_mishra_patait_2023}. Experiments were distributed across multiple dedicated nodes to prevent physical reconfiguration variance; however, we enforced rigorous software consistency across all systems. All nodes operated within a controlled Microsoft Windows environment running the identical NVIDIA Display Driver version and FFmpeg build to ensure consistency. Detailed specifications for each compute node are available in the \textit{Supplementary Material}.

\vspace{-3mm}

\subsection{Encoding Throughput Benchmark} \label{sec_thpt}

To isolate the peak encoding throughput of NVENC and eliminate other potential bottlenecks, such as PCIe bandwidth, disk I/O latency, or system memory (DRAM) throughput, we utilized a synthetic frame generation methodology. Following protocols established in our previous work \cite{11417632}, we utilized FFmpeg's \textit{lavfi} source filter to generate a black frame, which was uploaded to the GPU VRAM via the \textit{hwupload} filter. This single reference frame was then looped internally within the GPU for the duration of the encoding session, which ensures that the measured frame rates represent the absolute upper bound of the NVENC hardware.

We evaluated HEVC performance across all four generations and two tuning modes, HQ and UHQ, except the Pascal GPU, which predates UHQ support. AV1 evaluation was performed only on Ada Lovelace and Blackwell architectures, as prior generations lack hardware support for this code. All tests were conducted using an 8-bit 4:2:0 pixel format to align with current industry standards for Standard Dynamic Range (SDR) live streaming. We benchmarked three distinct preset configurations,P1 (Fastest), P4 (Medium), and P7 (Slowest), at a target resolution of 4K UHD ($3840\times2160$) with a constant bitrate (CBR) of 30 Mbps. Each test run processed a duration of 2000 frames to ensure thermal steady-state, after which the average encoding speed was recorded. Detailed encoding parameters are defined in \cref{Sec_RD}, and the benchmark scripts are provided in the \textit{Supplementary Material}.
\vspace{-5mm}
\subsection{Power Consumption and GPU Utilization Analysis}
To assess the chip utilization and power characteristics of the GPU under sustained encoding loads, we monitored the Board Power Draw, GPU Chip Power Draw, and GPU Load during the encoding sessions. Telemetry data was logged using GPU-Z (v2.68.0). For each configuration, the encoding workload was sustained for a fixed duration of 30 seconds to negate initial boost-clock transients and ensure the GPU reached a representative thermal steady state. Telemetry was logged at 1-second intervals, from which the average of each parameter was calculated and reported.
\vspace{-3mm}
\subsection{Rate-Distortion Analysis} \label{Sec_RD}
\vspace{-0.5mm}
To evaluate the compression efficiency of the NVENC hardware across diverse content characteristics, we adopted a dual-dataset methodology mirroring the vendor's validation protocols \cite{nvidia_bw_2024, nvidia_2024b, nvidia_2024}. The experiment consists of two distinct content categories: Natural Content and Gaming Content.

For Natural Content, we utilized \textit{Netflix Chimera} dataset \cite{netflix}, a DCI 4K ($4096\times2160$) sequence at 59.94 fps, featuring 23 distinct scenes with a total runtime of 30:49 minutes. As the source material was provided as an HDR TIFF sequence in the Display P3 color space, we performed a high-fidelity conversion to SDR and center-cropped the frame to a standard UHD resolution of $3840\times2160$. To accelerate the encoding pipeline, we transcoded these sequences into a visually lossless mezzanine format using x265 with CRF of 10, enabling hardware-accelerated decoding via the NVDEC. On the other hand, for Gaming Content, we utilized the Twitch Dataset, as available on \textit{xiph.org}, featuring ten distinct one-minute sequences of the ten most popular games streamed on Twitch ($1920\times1080$ at 60 fps) \cite{xiph_org}. The original source files, provided in Y4M format (RGB), were converted to the standard 4:2:0 YUV color space with 8-bit depth. Following established methodology in our prior work \cite{10637525}, we upscaled these sequences to 4K UHD using a Lanczos filter before encoding them to the same mezzanine format to maintain consistency across the test bench. We evaluated the same matrix of codec/preset combinations defined in \cref{sec_thpt}, adhering to generational hardware limitations. To strictly enforce real-time broadcast requirements, we enforced the following encoding parameters. Note that while manual control over B-frames (-bf) and reference frames (-refs) was applied to the HQ tuning mode, these flags were omitted for UHQ tests as the mode prevented manual user configuration.

\begin{itemize}
\item \textbf{Rate Control:} Constant Bitrate (CBR) with a Video Buffer Verifier (VBV) buffer size set to $2\times$ the target bitrate.
\item \textbf{GOP Structure:} Fixed 2-second Group of Pictures (closed GOP).
\item \textbf{Frame Structure (HQ):} 2 B-frames and 1 Reference frame (\textit{-bf 2 -refs 1}).
\item \textbf{Format:} HEVC (H.265) / AV1 Main profile, 4:2:0 chroma subsampling, 8-bit depth.
\item \textbf{Optimization:} Spatial and Temporal Adaptive Quantization were disabled (\textit{-spatial\_aq 0 -temporal\_aq 0}). Lookahead was disabled (\textit{-rc-lookahead 0}) to minimize latency. Split Frame Encoding (SFE) was disabled (\textit{-split\_encode\_mode 15}) to ensure single-engine performance consistency.
\end{itemize}

Rate-distortion performance was quantified using the Bjøntegaard Delta Rate (BD-Rate) metric \cite{barman2024bjontegaarddeltabdtutorial}, derived from ten distinct bitrate points (1, 2, 3, 4, 7, 10, 15, 22, 35, 50 Mbps) to ensure curve accuracy. We calculated both PSNR-Y and VMAF metrics using the \textit{vmaf\_cuda} library on the dedicated system detailed in \cref{tab:hardware}. For VMAF, we utilized the \textit{vmaf\_4k\_v0.6.1neg} model, which is optimized to isolate pure compression artifacts \cite{li_swanson_bampis_krasula_aaron_2020}. While VMAF scores are provided in the \textit{Supplementary Material} for reference, this paper primarily reports PSNR-Y BD-Rate values to maintain consistency with standard signal-processing metrics. Furthermore, to streamline the analysis, the main text reports only the aggregated average BD-Rate for each dataset; a granular breakdown of per-sequence performance is available in the \textit{Supplementary Material}.

\vspace{-3mm}
\subsection{GOP Structure Analysis}

To analyze the temporal characteristics of the encoded bitstreams, specifically the B-frame hierarchy and reference frame utilization that are obscured by the UHQ tuning mode, we employed a frame-level inspection methodology using FFmpeg's bitstream filters and probe tools. For HEVC, we determined the B-frame depth by explicitly extracting the \textit{pict\_type} entries for the initial sequence intervals. To quantify the active reference frame count, a critical factor for decoder compatibility and latency, we utilized the \textit{trace\_headers} filter to parse the \textit{num\_ref\_idx\_l0} (List 0 reference index) syntax elements from the slice headers of P-frames. 

For AV1, given that the architecture maintains a fixed reference frame pool of 8, our analysis focused on deriving the effective B-frame structure. We extracted the \textit{order\_hint} parameters from the sequence headers by passing the bitstream through the \textit{trace\_headers} filter. By analyzing the delta between the presentation order and the decoding order hints over the initial 30 frames, we reconstructed the hierarchical B-frame pattern employed by the encoder. This allowed us to explicitly verify whether the UHQ mode increases the reference pool size compared to the standard HQ configuration.
\vspace{-7mm}
\subsection{End-to-End Latency Measurement}

Following the methodology in our recent work \cite{11396901}, we evaluated E2E latency using the Open Broadcaster Software (OBS). A source video with a 60 fps timecode was streamed at 30 Mbps CBR from OBS via the WHIP protocol to a local SRS server, and received by a client via WHEP on the same local network. Latency was measured by calculating the timecode difference between the source and playback monitors captured in high-shutter-speed photographs. All latency measurements were conducted on the Ada Lovelace systems (Encoder: RTX 4070 Ti SUPER, Decoder: RTX 2000 Ada Laptop) to minimize hardware variance. The specification of the Encoding and Decoding system can be found in \textit{Supplementary Material}.
\vspace{-3.5mm}
\section{Results and Analysis}
\vspace{-1.5mm}
\subsection{Encoding Throughput} \label{sec_EncodeThpt}

\begin{table}[!tbp]
\vspace{-2mm}
\setstretch{0.8}
\vspace{-0.5mm}
\caption{4K Encoding Throughput (FPS) by Configuration}
\centering
\label{tab:EncodingThpt}
\resizebox{5cm}{!}{\begin{tabular}{@{}lcccccc@{}}
\toprule
\multirow{2.5}{*}{Architecture}& \multicolumn{3}{c}{HQ} & \multicolumn{3}{c}{UHQ} \\
\cmidrule(lr){2-4} \cmidrule(lr){5-7}
& P1 & P4 & P7 & P1 & P4 & P7 \\
\midrule
\multicolumn{7}{c}{H.265/HEVC}\\\midrule
Pascal & 123&110&108&N/A&N/A&N/A\\
Ampere & 150&109&39&50&38&17\\
Ada Lovelace & 288&128&46&101&74&34\\
Blackwell & 315&142&45&104&80&37\\
\midrule
\multicolumn{7}{c}{AV1}\\\midrule
Ada Lovelace & 212&126&86&89&63&36\\
Blackwell & 256&138&104&107&74&46\\
\bottomrule
\end{tabular}}
\vspace{-4mm}
\end{table}

\cref{tab:EncodingThpt} illustrates the severe computational penalty associated with UHQ tuning, confirming that the processing pipeline used in this configuration acts as a significant throttle on the encoding throughput. While the Blackwell architecture demonstrates a robust generational uplift, achieving a peak HEVC throughput of 315 FPS compared to Ada Lovelace's 288 FPS in the P1 preset, the activation of UHQ mode poses a significant impact, resulting in a drastic decline in performance, particularly for AV1. For Blackwell AV1 encoding at the P7 preset, the throughput dropped from 104 FPS (HQ) to 46 FPS (UHQ), a reduction of approximately 56\%, effectively pushing the throughput dangerously close to the 30 FPS real-time threshold. A similar trend is observed in the Ada Lovelace architecture, where AV1 P7 throughput drops from 86 FPS to 36 FPS. Notably, for HEVC at the P7 preset, the throughput delta between HQ (45 FPS) vs UHQ (37 FPS) on Blackwell is less pronounced than in AV1, likely because the HEVC P7 implementation is already heavily bottlenecked by complex mode decisions even in HQ mode. However, across all modern architectures, UHQ tuning consistently reduces encoding speed by 50–60\% compared to equivalent HQ presets, validating the hypothesis that offloading tasks to CUDA cores introduces a substantial serialization bottleneck as frames needed to be processed by the GPU Core.
\vspace{-3mm}
\subsection{Power Consumption and GPU Utilization}
\vspace{-1mm}
\begin{table}[!tbp]
\vspace{-2mm}
\setstretch{0.80}
\vspace{-0.5mm}
\caption{Main GPU Core Utilization (\%) by Configuration}
\centering
\label{tab:GPUUsage}
\resizebox{5cm}{!}{\begin{tabular}{@{}lcccccc@{}}
\toprule
\multirow{2.5}{*}{Architecture}& \multicolumn{3}{c}{HQ} & \multicolumn{3}{c}{UHQ} \\
\cmidrule(lr){2-4} \cmidrule(lr){5-7}
& P1 & P4 & P7 & P1 & P4 & P7 \\
\midrule
\multicolumn{7}{c}{H.265/HEVC}\\\midrule
Pascal & 2&2&2&N/A&N/A&N/A\\
Ampere & 2&1&0&18&13&7\\
Ada Lovelace & 1&1&0&10&7&3\\
Blackwell & 1&1&0&15&12&6\\
\midrule
\multicolumn{7}{c}{AV1}\\\midrule
Ada Lovelace & 1&1&0&8&6&4\\
Blackwell & 1&1&0&13&9&6\\
\bottomrule
\end{tabular}}
\vspace{-6mm}
\end{table}

The telemetry data in \cref{tab:GPUUsage} and \cref{tab:PowerConsumption} reveal a cost to the hybrid approach utilized by UHQ tuning. Unlike the standard HQ mode, which operates almost exclusively on the fixed-function SIP block (registering negligible 0-2\% compute utilization), the UHQ mode introduces a non-trivial load on the general-purpose SM. For instance, on the Blackwell GPU, AV1 UHQ encoding at the P1 preset consumes up to 13\% of the main GPU core resources, while Ampere HEVC UHQ peaks at 18\%. This resource contention presents a critical bottleneck for single-PC live streaming setups, where the encoder must compete for the same SM required by the game engine, potentially degrading in-game frame rate. Furthermore, the power data in \cref{tab:PowerConsumption} demonstrates the inefficiency of this hybrid approach, as enabling UHQ on Ada Lovelace (HEVC P1) spikes board power consumption from 52.2 W to 72.5 W, a nearly 40\% increase. While the Blackwell architecture appears to mitigate this penalty through improved architectural design, maintaining a relatively flat power delta between HQ and UHQ, the preceding generations incur a severe energy cost, making UHQ tuning a suboptimal choice for power-constrained high-density deployment scenarios. \looseness=-1
\begin{table}[!tbp]
\vspace{-2mm}
\setstretch{0.75}
\vspace{-0.5mm}
\caption{Power Consumption (W) by Configuration}
\centering
\label{tab:PowerConsumption}
\resizebox{5.5cm}{!}{\begin{tabular}{@{}lcccccc@{}}
\toprule
\multirow{2.5}{*}{Architecture}& \multicolumn{3}{c}{HQ} & \multicolumn{3}{c}{UHQ} \\
\cmidrule(lr){2-4} \cmidrule(lr){5-7}
& P1 & P4 & P7 & P1 & P4 & P7 \\
\midrule
\multicolumn{7}{c}{\textbf{Board Power Consumption (Including VRAM)}}\\\midrule
\multicolumn{7}{c}{H.265/HEVC}\\\midrule
Pascal&49.7&49.5&50.4&N/A&N/A&N/A      \\
Ampere&53.5&54.3&53.7&64.2&64.1&55.2   \\
Ada Lovelace&52.2&52.6&50.3&72.5&68.8&59.8      \\
Blackwell&55.1&53.4&51.1&62.9&68.8&60.1\\
\midrule
\multicolumn{7}{c}{AV1}\\\midrule
Ada Lovelace&54.8&53.5&58.2&65.7&59.9&56.3      \\
Blackwell&57.8&55.3&54.4&66.8&65&62.7  \\
\midrule
\multicolumn{7}{c}{\textbf{GPU Chip Power Consumption}}\\\midrule
\multicolumn{7}{c}{H.265/HEVC}\\\midrule
Pascal&21.9&22.2&22.7&N/A&N/A&N/A       \\
Ampere&31.8&32.9&33.7&41.3&41.5&33.5    \\
Ada Lovelace&35.2&35.6&34.1&55&51.8&43  \\
Blackwell&21.2&20.6&19.6&24.2&26.2&23.2 \\
\midrule
\multicolumn{7}{c}{AV1}\\\midrule
Ada Lovelace&37.3&36.5&36&48.5&42.2&38.9\\
Blackwell&22.2&21.2&20.9&25.8&25&24.1   \\
\bottomrule
\end{tabular}}
\vspace{-3mm}
\end{table}
\vspace{-1mm}
\subsection{Rate-Distortion (RD)} \label{sec_RD}
\vspace{-1mm}
\begin{table}[!tbp]
\vspace{-2mm}
\setstretch{0.76}
\vspace{-0.5mm}
\caption{BD-Rate PSNR-Y (\%) (Pascal HQ [HEVC] Anchor)}
\vspace{0.5mm}
\centering
\label{tab:PascalHQBDRate}
\resizebox{8.5cm}{!}{\begin{tabular}{@{}lcccccccccc@{}}
\toprule
\multirow{2.5}{*}{Architecture}& \multirow{2.5}{*}{Tune}& \multicolumn{3}{c}{Netflix Chimera} & \multicolumn{3}{c}{Twitch} & \multicolumn{3}{c}{Average} \\
\cmidrule(lr){3-5} \cmidrule(lr){6-8} \cmidrule(lr){9-11}
&& P1 & P4 & P7 & P1 & P4 & P7 & P1 & P4 & P7\\
\midrule
\multirow{2}{*}{Ampere}
&HQ&-12.10&-7.04&-6.54&-6.67&-2.46&-1.55&-9.38&-4.75&-4.05          \\
&UHQ&-22.68&-19.10&-19.36&-14.87&-11.41&-12.19&-18.77&-15.25&-15.77 \\
\midrule
\multirow{2}{*}{Ada Lovelace}
&HQ&-11.64&-6.50&-6.64&-6.60&-2.15&-1.53&-9.12&-4.33&-4.09          \\
&UHQ&-22.37&-18.63&-18.63&-14.83&-11.33&-12.05&-18.60&-14.98&-15.34 \\
\midrule
\multirow{2}{*}{Blackwell}
&HQ&-12.96&-10.96&-11.22&-9.31&-8.46&-7.93&-11.13&-9.71&-9.57       \\
&UHQ&-24.46&-25.98&-25.96&-18.41&-19.35&-19.62&-21.43&-22.66&-22.79 \\
\bottomrule
\end{tabular}}
\vspace{-4mm}
\end{table}

\begin{table}[!tbp]
\vspace{-2mm}
\setstretch{0.76}
\vspace{-0.5mm}
\caption{BD-Rate PSNR-Y (\%) (Ada Lovelace HQ Anchor)}
\vspace{0.5mm}
\centering
\label{tab:AdaHQBDRate}
\resizebox{8.5cm}{!}{\begin{tabular}{@{}lcccccccccc@{}}
\toprule
\multirow{2.5}{*}{Architecture}& \multirow{2.5}{*}{Tune}& \multicolumn{3}{c}{Netflix Chimera} & \multicolumn{3}{c}{Twitch} & \multicolumn{3}{c}{Average} \\
\cmidrule(lr){3-5} \cmidrule(lr){6-8} \cmidrule(lr){9-11}
&& P1 & P4 & P7 & P1 & P4 & P7 & P1 & P4 & P7\\
\midrule
\multicolumn{11}{c}{H.265/HEVC}\\\midrule

\multirow{2}{*}{Ampere}
&HQ&-0.54&-0.56&0.02&-0.08&-0.33&-0.02&-0.31&-0.45&0.00            \\
&UHQ&-12.79&-13.50&-13.75&-7.20&-7.94&-9.72&-10.00&-10.72&-11.74   \\
\midrule

\multirow{2}{*}{Blackwell}
&HQ&-1.51&-5.28&-5.31&-2.97&-6.52&-6.57&-2.24&-5.90&-5.94          \\
&UHQ&-13.38&-18.32&-18.43&-9.82&-14.64&-15.76&-11.60&-16.48&-17.09 \\
\midrule
\multicolumn{11}{c}{AV1}\\\midrule
\multirow{2}{*}{Blackwell}
&HQ&1.02&2.11&2.11&-2.04&-1.96&-1.19&-0.51&0.07&0.46               \\
&UHQ&-18.30&-18.67&-19.17&-20.30&-21.40&-20.66&-19.30&-20.03&-19.91\\
\bottomrule
\end{tabular}}
\vspace{-4mm}
\end{table}

\begin{table}[!tbp]
\vspace{-2mm}
\setstretch{0.76}
\vspace{-0.5mm}
\caption{BD-Rate PSNR-Y (\%) (HQ vs. UHQ)}
\vspace{0.5mm}
\centering
\label{tab:HQvsUHQBDRate}
\resizebox{8.5cm}{!}{\begin{tabular}{@{}lccccccccc@{}}
\toprule
\multirow{2.5}{*}{Architecture}& \multicolumn{3}{c}{Netflix Chimera} & \multicolumn{3}{c}{Twitch} & \multicolumn{3}{c}{Average} \\
\cmidrule(lr){2-4} \cmidrule(lr){5-7} \cmidrule(lr){8-10}
& P1 & P4 & P7 & P1 & P4 & P7 & P1 & P4 & P7\\
\midrule
\multicolumn{10}{c}{H.265/HEVC}\\\midrule

Ampere&-12.37&-13.05&-13.87&-9.12&-9.51&-11.18&-10.74&-11.28&-12.53        \\
Ada Lovelace&-12.49&-13.04&-12.96&-9.16&-9.69&-11.08&-10.83&-11.37&-12.02  \\
Blackwell&-11.99&-13.24&-13.39&-10.26&-12.03&-12.87&-11.13&-12.63&-13.13   \\\midrule
\multicolumn{10}{c}{AV1}\\\midrule
Ada Lovelace&-15.62&-16.52&-16.75&-13.61&-14.92&-14.95&-14.62&-15.72&-15.85\\
Blackwell&-18.58&-19.70&-20.26&-15.50&-16.37&-16.90&-17.04&-18.03&-18.58   \\
\bottomrule
\end{tabular}}
\vspace{-4mm}
\end{table}

\begin{table}[!tbp]
\vspace{-2mm}
\setstretch{0.76}
\vspace{-0.5mm}
\caption{BD-Rate PSNR-Y (\%) (Ada Lovelace UHQ Anchor)}
\vspace{0.5mm}
\centering
\label{tab:AdaUHQBDRate}
\resizebox{8.5cm}{!}{\begin{tabular}{@{}lccccccccc@{}}
\toprule
\multirow{2.5}{*}{Architecture}& \multicolumn{3}{c}{Netflix Chimera} & \multicolumn{3}{c}{Twitch} & \multicolumn{3}{c}{Average} \\
\cmidrule(lr){2-4} \cmidrule(lr){5-7} \cmidrule(lr){8-10}
& P1 & P4 & P7 & P1 & P4 & P7 & P1 & P4 & P7\\
\midrule
\multicolumn{10}{c}{H.265/HEVC}\\\midrule
Ampere&-0.07&-0.46&-1.22&2.90&2.71&2.34&1.42&1.12&0.56            \\
Blackwell&-4.42&-10.70&-10.23&-4.03&-8.61&-8.23&-4.22&-9.66&-9.23 \\\midrule
\multicolumn{10}{c}{AV1}\\\midrule
Blackwell&2.52&3.02&2.54&-4.22&-3.84&-3.62&-0.85&-0.41&-0.54      \\
\bottomrule
\end{tabular}}
\vspace{-6mm}
\end{table}

The longitudinal analysis against the legacy Pascal baseline (\cref{tab:PascalHQBDRate}) reveals a non-linear evolution in HEVC coding efficiency. While the transition from Pascal to Ampere and Ada Lovelace yielded stagnant returns in standard HQ mode, hovering around a 4\% average BD-Rate saving, the Blackwell architecture breaks this plateau. In standard HQ mode, Blackwell achieves a 5.94\% BD-Rate saving over Ada Lovelace (\cref{tab:AdaHQBDRate}) (Preset P7), a result that closely aligns with the 5\% improvement cited in NVIDIA’s whitepaper \cite{nvidia_bw_2024, nvidia_2024b}. This generational leap, which effectively doubles the incremental gains seen in previous iterations, is potentially attributable to the introduction of native 10-bit internal processing for HEVC, which allows for higher-precision mode decisions and residual coding even when processing 8-bit input sources.

\cref{tab:HQvsUHQBDRate} isolates the impact of the hybrid encoding model, demonstrating that UHQ consistently delivers double-digit bitrate reductions, averaging 12–13\% for HEVC and 16–19\% for AV1 (Preset P7), albeit at the throughput costs detailed in \cref{sec_EncodeThpt}. For AV1, while the standard HQ performance is statistically identical between Ada and Blackwell (+0.46\% BD-Rate, \cref{tab:AdaHQBDRate}), the Blackwell architecture is significantly more aggressive in its UHQ implementation, achieving an 18.58\% reduction against its own HQ baseline compared to Ada's 15.85\% reduction. This suggests that while fixed-function AV1 performance has plateaued, the hybrid compute-offload capabilities in Blackwell are more efficient at extracting temporal redundancy. BD-Rate VMAF data in \textit{Supplementary Material} reveals that while UHQ excels in complex scenes, it can perceptually regress in simpler, low-motion sequences. A full investigation into these content-dependent failure cases is reserved for future work.

\vspace{-3mm}
\subsection{GOP Structure}

\begin{table}[!tbp]
\vspace{-2mm}
\setstretch{0.76}
\vspace{-0.5mm}
\caption{Active Reference Frame and B-Frame Count}
\vspace{0.5mm}
\centering
\label{tab:GOPStructure}
\resizebox{5cm}{!}{\begin{tabular}{@{}lcccccc@{}}
\toprule
\multirow{2.5}{*}{Architecture}& \multicolumn{3}{c}{HQ} & \multicolumn{3}{c}{UHQ} \\
\cmidrule(lr){2-4} \cmidrule(lr){5-7}
& P1 & P4 & P7 & P1 & P4 & P7 \\
\midrule
\multicolumn{7}{c}{\textbf{Active Reference Frame Count}}\\\midrule
\multicolumn{7}{c}{H.265/HEVC}\\\midrule
Pascal & 1&1&1&N/A&N/A&N/A\\
Ampere & 1&1&1&1&2&4\\
Ada Lovelace & 1&1&1&1&2&4\\
Blackwell & 1&1&1&1&2&4\\
\midrule
\multicolumn{7}{c}{\textbf{B-Frame Count}}\\\midrule
\multicolumn{7}{c}{H.265/HEVC}\\\midrule
Pascal & 0&0&0&N/A&N/A&N/A\\
Ampere & 2&2&2&5&5&5\\
Ada Lovelace & 2&2&2&5&5&5\\
Blackwell & 2&2&2&5&5&5\\
\midrule
\multicolumn{7}{c}{AV1}\\\midrule
Ada Lovelace & 2&2&2&5&5&5\\
Blackwell & 2&2&2&7&7&7\\
\bottomrule
\end{tabular}}
\vspace{-4mm}
\end{table}
\begin{table}[!tbp]
\vspace{-2mm}
\setstretch{0.76}
\vspace{-0.5mm}
\caption{Measured End-to-End latency in number of frames on 60p Timecode. `SFE' = SFE was used. `DNR' = `Did Not Run' or failed to achieve real-time encoding even with SFE.}
\vspace{0.5mm}
\centering
\label{tab:E2ELatency}
\resizebox{6cm}{!}{\begin{tabular}{@{}lcccccc@{}}
\toprule
\multirow{2.5}{*}{Codec}& \multicolumn{3}{c}{HQ} & \multicolumn{3}{c}{UHQ} \\
\cmidrule(lr){2-4} \cmidrule(lr){5-7}
& P1 & P4 & P7 & P1 & P4 & P7 \\
\midrule
H.265/HEVC&7&7&DNR&20&47 (SFE)&DNR\\
AV1&7&7&7&38&52 (SFE)&DNR\\
\bottomrule
\end{tabular}}
\vspace{-6mm}
\end{table}

\cref{tab:GOPStructure} reveals the aggressive temporal restructuring concealed by the UHQ Tuning. While the standard HQ tuning maintains a conservative, low-latency profile with a single active reference frame and a shallow hierarchy of 2 B-frames (post-Pascal), as configured, activating UHQ mode forces the encoder to drastically expand the Group of Pictures (GOP). For HEVC, all modern architectures immediately lock to a fixed 5 B-frame structure across all presets, but the most significant divergence appears in AV1: while Ada Lovelace caps UHQ at 5 B-frames, the Blackwell architecture pushes this depth to 7 B-frames, the deepest hierarchy observed in any consumer hardware encoder to date. Furthermore, at the highest quality preset (P7), HEVC UHQ quadruples the active reference frame count from 1 to 4. Note that active reference frame counts for AV1 are excluded, as the AV1 specification mandates a fixed pool of 8 reference frames. This mandatory expansion of the Decoded Picture Buffer (DPB) provides the necessary temporal search space to achieve the $\approx$18\% BD-Rate gains noted in \cref{sec_RD}, but it does so by physically enforcing a multi-frame buffer debt that prevents ultra-low-latency transmission and potentially affects compatibility.
\vspace{-7mm}
\subsection{End-to-End Latency}

\cref{tab:E2ELatency} demonstrates the prohibitive latency penalty of UHQ tuning, effectively disqualifying it for interactive applications despite its coding efficiency. While standard HQ mode maintains a consistent, low-latency baseline of 7 frames ($\approx$117 ms) across amenable configurations, UHQ mode introduces a massive serialization delay, spiking HEVC P1 latency to 20 frames and AV1 P1 to 38 frames. The computational burden at the P4 preset is severe enough to necessitate SFE merely to sustain sufficient throughput, resulting in a latency of 47–52 frames, while the highest quality P7 preset fails to achieve real-time throughput entirely (DNR). Consequently, UHQ tuning violates the strict motion-to-photon thresholds required for cloud gaming, relegating its utility strictly to non-interactive broadcast or archival workflows where sub-second latency is not a critical constraint.
\vspace{-3mm}
\section{Conclusions and Future Work}
\vspace{-2mm}
In this paper, we conducted a longitudinal performance analysis of NVIDIA NVENC, tracing the evolution from Pascal to Blackwell to determine the viability of emerging hybrid encoding modes for real-time applications. Moreover, we investigated the newly introduced UHQ Tuning, which promised to significantly improve the compression efficiency. Our evaluation confirms that while Blackwell breaks historical efficiency plateaus, achieving a 5.94\% BD-Rate gain in standard modes and up to 22.79\% in UHQ modes, these gains are not free. By deconstructing the UHQ abstraction, we revealed that the mode relies on a brute-force expansion of the temporal search space, utilizing up to 7 B-frames and quadrupled reference counts. This architectural shift fundamentally alters the hardware encoder's profile, trading the sub-frame latency required for cloud gaming for the high-delay, high-efficiency compression typical of archival transcoding, thereby disqualifying UHQ for interactive use cases.

Future work will pivot towards reconciling these high-efficiency capabilities with the strict constraints of ultra-low-latency streaming. Specifically, we aim to investigate intermediate tuning configurations that leverage Blackwell's native 10-bit processing and architectural improvements without incurring the prohibitive buffering depth of the full UHQ implementation. Furthermore, we plan to extend this analysis to 5G network field tests, evaluating how dynamically adjusting these temporal structures can optimize transmission within the rigid uplink constraint of commercial TDD networks. Such research is essential for developing adaptive rate-control algorithms capable of sustaining high-fidelity volumetric streaming over fluctuating cellular uplinks.

\vspace{-1.5mm}

\setstretch{0.94}
\newcommand{\BIBdecl}{\setlength{\itemsep}{0.1 em}}
\bibliographystyle{IEEEtran}
\bibliography{refs}

\newpage

\appendix

\section{Hardware and Software Configuration}

\begin{table}[!h]
\setstretch{0.8}
\caption{Hardware and Software Configuration}
\centering
\label{tab:hardware}
\resizebox{7.7cm}{!}{\begin{tabular}{@{}ll@{}}
\toprule
\multicolumn{2}{c}{\textbf{Pascal Encoding System}}\\
\midrule
Hardware                 & Description  \\\midrule
System Board & ASUS PRIME B365M-A\\
CPU & Intel(R) Xeon(R) E-2224 CPU @ 3.40GHz \\
RAM & Dual-Channel DDR4 20 GB @ 2400 MT/s \\ 
NVIDIA GPU & NVIDIA GeForce GTX 1070\\\midrule
Software & Version \\\midrule
OS & Microsoft Windows 11 Pro Build 22631\\
ffmpeg & N-122268-g0dfaed77a6\\
NVIDIA GPU Driver & GeForce Game Ready Driver 581.80 (581.80)\\
\midrule
\multicolumn{2}{c}{\textbf{Ampere Encoding System}}\\
\midrule
Hardware                 & Description  \\\midrule
System Board & ASRock B660M-ITX\\
CPU & 12th Gen Intel(R) Core(TM) i5-12400F \\
RAM & Dual-Channel DDR4 64 GB @ 3200 MT/s \\ 
NVIDIA GPU & NVIDIA GeForce RTX 3060 12 GB (GA104)\\\midrule
Software & Version \\\midrule
OS & Microsoft Windows 11 Pro Build 26200\\
ffmpeg & N-122268-g0dfaed77a6\\
NVIDIA GPU Driver & GeForce Game Ready Driver 581.80 (581.80)\\\midrule
\multicolumn{2}{c}{\textbf{Ada Lovelace Encoding System}}\\
\midrule
Hardware                 & Description  \\\midrule
System Board & ASUS TUF GAMING Z890-PRO WIFI\\
CPU & Intel(R) Core(TM) Ultra 9 285K \\
RAM & Dual-Channel DDR5 128 GB @ 6000 MT/s \\ 
NVIDIA GPU & NVIDIA GeForce RTX 4070 Ti SUPER\\\midrule
Software & Version \\\midrule
OS & Microsoft Windows Server 2025 Datacenter Build 26100\\
ffmpeg & N-122268-g0dfaed77a6\\
NVIDIA GPU Driver & GeForce Game Ready Driver 581.80 (581.80)\\\midrule
\multicolumn{2}{c}{\textbf{Blackwell Encoding System}}\\
\midrule
Hardware                 & Description  \\\midrule
System Board & MSI Z390-S01\\
CPU & Intel(R) Core(TM) i7-8700K CPU @ 3.70GHz \\
RAM & Dual-Channel DDR4 64 GB @ 2666 MT/s \\ 
NVIDIA GPU & NVIDIA GeForce RTX 5070 Ti\\\midrule
Software & Version \\\midrule
OS & Microsoft Windows 10 Pro Build 19045\\
ffmpeg & N-122268-g0dfaed77a6\\
NVIDIA GPU Driver & GeForce Game Ready Driver 581.80 (581.80)\\
\bottomrule
\end{tabular}}
\vspace{-7mm}
\end{table}

\begin{table}[!h]
\setstretch{0.8}
\caption{Hardware and Software Configuration (Continue)}
\centering
\label{tab:hardware2}
\resizebox{7.7cm}{!}{\begin{tabular}{@{}ll@{}}
\toprule
\multicolumn{2}{c}{\textbf{End-to-End Latency Encoding System}}\\
\midrule
Hardware                 & Description  \\\midrule
System Board & ASUS ROG STRIX X570-F GAMING\\
CPU & AMD Ryzen 9 5900X 12-Core Processor \\
RAM & Dual-Channel DDR4 96 GB @ 3600 MT/s \\ 
NVIDIA GPU & NVIDIA GeForce RTX 4070 Ti SUPER\\\midrule
Software & Version \\\midrule
OS & Microsoft Windows 10 Pro Build 19045\\
NVIDIA GPU Driver & GeForce Game Ready Driver 577.00 (577.00)\\
OBS Studio & 32.0.4\\
\midrule
\multicolumn{2}{c}{\textbf{End-to-End Latency Decoding System}}\\
\midrule
Hardware                 & Description  \\\midrule
System Board & Dell Precision 5680\\
CPU & 13th Gen Intel(R) Core(TM) i7-13700H \\
RAM & Octal-Channel LPDDR5x 32 GB @ 6000 MT/s \\ 
NVIDIA GPU & NVIDIA RTX 2000 Ada Generation Laptop\\\midrule
Software & Version \\\midrule
OS & Microsoft Windows 11 Pro Build 22631\\
NVIDIA GPU Driver & GeForce Game Ready Driver 577.00 (577.00)\\
\midrule

\multicolumn{2}{c}{\textbf{WebRTC Server}}\\
\midrule
Hardware                 & Description  \\\midrule
System Board & Dell PowerEdge R7515\\
CPU & AMD EPYC 7C13 64-Core Processor\\
RAM & Octal-Channel DDR4 512 GB @ 2666 MT/s \\\midrule
Software & Version \\\midrule
OS & Ubuntu 24.04.1 LTS\\
WebRTC Server & SRS 6.0.101\\\midrule
\multicolumn{2}{c}{\textbf{VMAF Calculation System}}\\
\midrule
Hardware                 & Description  \\\midrule
CPU & Intel(R) Core(TM) Ultra 9 285K \\
RAM & Dual-Channel DDR5 128 GB (4×32 GB) @ 4400 MT/s \\ 
GPU & NVIDIA RTX PRO 5000 48GB Blackwell\\\midrule
Software & Version \\\midrule
OS & Ubuntu 24.04.3 LTS\\
ffmpeg & N-122271-g0629780cf6\\
libvmaf & v3.0.0 (b9ac69e6)\\
VMAF Model & vmaf\_4k\_v0.6.1neg \\
NVIDIA GPU Driver & 580.126.09 \\
NVIDIA CUDA Compiler & cuda\_13.1.r13.1\/compiler.36836380\_0 \\
\bottomrule
\end{tabular}}
\vspace{-7mm}
\end{table}

\clearpage
\onecolumn
\section{Encoding Throughput Benchmark Script}

\subsection{HEVC 8-Bit 4:2:0 (HQ)}

Please note that the flags \textit{-bf 2 -refs 1 -b\_ref\_mode middle} were omitted from the Pascal GPU evaluation, as the architecture lacks the necessary hardware support for these temporal configurations. Retaining these parameters would result in immediate initialization failures; therefore, the encoder was allowed to default to its native capabilities for this specific generation.

\begin{lstlisting}[language=batch]
@echo off
setlocal enabledelayedexpansion

:: Output log file
set "LOGFILE=nvenc_ram_test-4K-HEVC-HQ.log"
set "RES=3840x2160"

echo NVENC RAM-Buffer Speed Test (No Decoder) Started > "%LOGFILE%"
echo Resolution: %RES% >> "%LOGFILE%"

:: =========================================================
:: TEST SET 1: 8-bit 4:2:0
:: =========================================================
echo. >> "%LOGFILE%"
echo ======================================================= >> "%LOGFILE%"
echo [TEST SET 1] 8-bit 4:2:0 (RAM Buffer -> PCIe -> NVENC) >> "%LOGFILE%"
echo ======================================================= >> "%LOGFILE%"

for %%P in (p1 p4 p7) do (
    for %%T in (hq) do (
        for %%S in (15) do (
            echo Testing 8-bit 4:2:0 - Preset: %%P - Tune: %%T - SFE: %%S
            echo [8-BIT] Preset: %%P Tune: %%T SFE: %%S >> "%LOGFILE%"
            
            ffmpeg.exe -hide_banner -y ^
            -init_hw_device cuda=cuda_dev:0 -filter_hw_device cuda_dev ^
            -f lavfi -i color=c=black:s=%RES%:r=60 ^
            -c:v hevc_nvenc -preset %%P -tune %%T ^
            -vf "format=yuv420p,trim=end_frame=1,hwupload=extra_hw_frames=64,
            loop=loop=-1:size=1:start=0" ^
            -spatial_aq 0 -temporal_aq 0 -rc-lookahead 0 -split_encode_mode %%S ^
            -profile:v main -b:v 30M -maxrate 30M -bufsize 60M -rc cbr ^
            -g 120 ^
            -frames:v 2000 ^
            -f null NUL 2>> "%LOGFILE%"
            
            echo --------------------------------------------------- >> "%LOGFILE%"
        )
    )
)


echo.
echo Tests Complete. Results saved in %LOGFILE%
pause
\end{lstlisting}

\newpage

\subsection{HEVC 8-Bit 4:2:0 (UHQ)}

\begin{lstlisting}[language=batch]
@echo off
setlocal enabledelayedexpansion

:: Output log file
set "LOGFILE=nvenc_ram_test-4K-HEVC-UHQ.log"
set "RES=3840x2160"

echo NVENC RAM-Buffer Speed Test (No Decoder) Started > "%LOGFILE%"
echo Resolution: %RES% >> "%LOGFILE%"

:: =========================================================
:: TEST SET 1: 8-bit 4:2:0
:: =========================================================
echo. >> "%LOGFILE%"
echo ======================================================= >> "%LOGFILE%"
echo [TEST SET 1] 8-bit 4:2:0 (RAM Buffer -> PCIe -> NVENC) >> "%LOGFILE%"
echo ======================================================= >> "%LOGFILE%"

for %%P in (p1 p4 p7) do (
    for %%T in (uhq) do (
        for %%S in (15) do (
            echo Testing 8-bit 4:2:0 - Preset: %%P - Tune: %%T - SFE: %%S
            echo [8-BIT] Preset: %%P Tune: %%T SFE: %%S >> "%LOGFILE%"
            
            ffmpeg.exe -hide_banner -y ^
            -init_hw_device cuda=cuda_dev:0 -filter_hw_device cuda_dev ^
            -f lavfi -i color=c=black:s=%RES%:r=60 ^
            -c:v hevc_nvenc -preset %%P -tune %%T ^
            -vf "format=yuv420p,trim=end_frame=1,hwupload=extra_hw_frames=64,
            loop=loop=-1:size=1:start=0" ^
            -spatial_aq 0 -temporal_aq 0 -rc-lookahead 0 -split_encode_mode %%S ^
            -profile:v main -b:v 30M -maxrate 30M -bufsize 60M -rc cbr ^
            -g 120 -b_ref_mode middle ^
            -frames:v 2000 ^
            -f null NUL 2>> "%LOGFILE%"
            
            echo --------------------------------------------------- >> "%LOGFILE%"
        )
    )
)


echo.
echo Tests Complete. Results saved in %LOGFILE%
pause
\end{lstlisting}

\subsection{AV1 8-Bit 4:2:0 (HQ)}
\begin{lstlisting}[language=batch]
@echo off
setlocal enabledelayedexpansion

:: Output log file
set "LOGFILE=nvenc_ram_test-4K-AV1_HQ.log"
set "RES=3840x2160"

echo NVENC RAM-Buffer Speed Test (No Decoder) Started > "%LOGFILE%"
echo Resolution: %RES% >> "%LOGFILE%"

:: =========================================================
:: TEST SET 1: 8-bit 4:2:0
:: =========================================================
echo. >> "%LOGFILE%"
echo ======================================================= >> "%LOGFILE%"
echo [TEST SET 1] 8-bit 4:2:0 (RAM Buffer -> PCIe -> NVENC) >> "%LOGFILE%"
echo ======================================================= >> "%LOGFILE%"

for %%P in (p1 p4 p7) do (
    for %%T in (hq) do (
        for %%S in (15) do (
            echo Testing 8-bit 4:2:0 - Preset: %%P - Tune: %%T - SFE: %%S
            echo [8-BIT] Preset: %%P Tune: %%T SFE: %%S >> "%LOGFILE%"
            
            ffmpeg.exe -hide_banner -y ^
            -init_hw_device cuda=cuda_dev:0 -filter_hw_device cuda_dev ^
            -f lavfi -i color=c=black:s=%RES%:r=60 ^
            -c:v av1_nvenc -preset %%P -tune %%T ^
            -vf "format=yuv420p,trim=end_frame=1,hwupload=extra_hw_frames=64,
            loop=loop=-1:size=1:start=0" ^
            -spatial_aq 0 -temporal_aq 0 -rc-lookahead 0 -split_encode_mode %%S ^
            -b:v 30M -maxrate 30M -bufsize 60M -rc cbr ^
            -g 120 -bf 2 -refs 1 -b_ref_mode middle ^
            -frames:v 2000 ^
            -f null NUL 2>> "%LOGFILE%"
            
            echo --------------------------------------------------- >> "%LOGFILE%"
        )
    )
)


echo.
echo Tests Complete. Results saved in %LOGFILE%
pause
\end{lstlisting}

\subsection{AV1 8-Bit 4:2:0 (UHQ)}
\begin{lstlisting}[language=batch]
@echo off
setlocal enabledelayedexpansion

:: Output log file
set "LOGFILE=nvenc_ram_test-4K-AV1_UHQ.log"
set "RES=3840x2160"

echo NVENC RAM-Buffer Speed Test (No Decoder) Started > "%LOGFILE%"
echo Resolution: %RES% >> "%LOGFILE%"

:: =========================================================
:: TEST SET 1: 8-bit 4:2:0
:: =========================================================
echo. >> "%LOGFILE%"
echo ======================================================= >> "%LOGFILE%"
echo [TEST SET 1] 8-bit 4:2:0 (RAM Buffer -> PCIe -> NVENC) >> "%LOGFILE%"
echo ======================================================= >> "%LOGFILE%"

for %%P in (p1 p4 p7) do (
    for %%T in (uhq) do (
        for %%S in (15) do (
            echo Testing 8-bit 4:2:0 - Preset: %%P - Tune: %%T - SFE: %%S
            echo [8-BIT] Preset: %%P Tune: %%T SFE: %%S >> "%LOGFILE%"
            
            ffmpeg.exe -hide_banner -y ^
            -init_hw_device cuda=cuda_dev:0 -filter_hw_device cuda_dev ^
            -f lavfi -i color=c=black:s=%RES%:r=60 ^
            -c:v av1_nvenc -preset %%P -tune %%T ^
            -vf "format=yuv420p,trim=end_frame=1,hwupload=extra_hw_frames=64,
            loop=loop=-1:size=1:start=0" ^
            -spatial_aq 0 -temporal_aq 0 -rc-lookahead 0 -split_encode_mode %%S ^
            -b:v 30M -maxrate 30M -bufsize 60M -rc cbr ^
            -g 120 -b_ref_mode middle ^
            -frames:v 2000 ^
            -f null NUL 2>> "%LOGFILE%"
            
            echo --------------------------------------------------- >> "%LOGFILE%"
        )
    )
)


echo.
echo Tests Complete. Results saved in %LOGFILE%
pause
\end{lstlisting}

\section{Rate-Distortion Benchmark Script}

The PowerShell script below is for encoding of the \textit{Twitch} dataset with HQ Tuning. For other datasets, replace the \textit{\$inputData} and \textit{\$bitrates} with the corresponding file names, frame rates, and desired bitrates. The AV1 codec has been removed for the evaluation on Pascal and Ampere GPUs, as the hardware is not supported. Similar to the Encoding Throughput Benchmark, flags \textit{-bf 2 -refs 1} were removed for the evaluation on the Pascal GPU. For UHQ Tuning, remove \textit{-bf 2 -refs 1} and change \textit{-tune hq} to \textit{-tune uhq}.

\begin{lstlisting}[language=PowerShell]
# --- CONFIGURATION ---

# 1. Concurrency Control
# Monitor VRAM. If "Out of Memory" occurs, lower this number.
$MaxConcurrentJobs = 2

# 2. Source Files & Framerates
$inputData = @(
    @{ Path="a01.mp4";  Fps=60000/1001 },
    @{ Path="a02.mp4";  Fps=60000/1001 },
    @{ Path="a03.mp4";  Fps=60000/1001 },
    @{ Path="a04.mp4";  Fps=60000/1001 },
    @{ Path="a05.mp4";  Fps=60000/1001 },
    @{ Path="a06.mp4";  Fps=60000/1001 },
    @{ Path="a07.mp4";  Fps=60000/1001 },
    @{ Path="a08.mp4";  Fps=60000/1001 },
    @{ Path="a09.mp4";  Fps=60000/1001 },
    @{ Path="a10.mp4";  Fps=60000/1001 }
)

# 3. Bitrate Ladder
$bitrates = 1, 2, 3, 4, 7, 10, 15, 22, 35, 50

# 4. Matrix Settings
$presets    = "p1", "p4", "p7"
$tunings    = "hevc", "av1"
$splitModes = @(
    @{ Name="1way"; Value=15 }
)

$baseOutputFolder = ''

# --- STEP 1: BUILD THE JOB QUEUE ---

$JobQueue = @()
New-Item -ItemType Directory -Force -Path $baseOutputFolder | Out-Null

Write-Host "Building Job Queue... This may take a moment." -ForegroundColor Cyan

foreach ($item in $inputData) {
    $file = $item.Path
    # Handle FPS calculation
    if ($item.Fps -is [string] -and $item.Fps -match "/") {
        $parts = $item.Fps -split "/"
        $fpsVal = [double]$parts[0] / [double]$parts[1]
    } else {
        $fpsVal = [double]$item.Fps
    }
    
    $gop = [math]::Round($fpsVal * 2)

    if (-not (Test-Path $file)) { continue }
    
    $fileNameWithoutExt = [System.IO.Path]::GetFileNameWithoutExtension($file)
    $sceneFolder = "$baseOutputFolder\$fileNameWithoutExt"
    if (-not (Test-Path $sceneFolder)) { New-Item -ItemType Directory -Force -Path $sceneFolder | Out-Null }

    foreach ($b in $bitrates) {
        $bufSize = $b * 2
        foreach ($preset in $presets) {
            foreach ($tune in $tunings) {
                foreach ($split in $splitModes) {
                    
                    $outputName = "$sceneFolder\${fileNameWithoutExt}_${b}Mbps_${preset}_${tune}.mp4"
                    
                    if (-not (Test-Path $outputName)) {
                        $JobQueue += [PSCustomObject]@{
                            InputFile  = $file
                            OutputFile = $outputName
                            Args       = @(
                                "-y",
                                "-hwaccel", "cuda", 
                                "-hwaccel_output_format", "cuda",
                                "-i", "`"$file`"",
                                "-vcodec", "${tune}_nvenc",
                                "-preset", "$preset",
                                "-tune", "hq",
                                "-spatial_aq", "0",
                                "-temporal_aq", "0",
                                "-rc-lookahead", "0",
                                "-split_encode_mode", "$($split.Value)",
                                "-b:v", "${b}M",
                                "-maxrate", "${b}M",
                                "-bufsize", "${bufSize}M",
                                "-rc", "cbr",
                                "-g", "$gop",
				                    "-bf", "2",
                                "-refs", "1",
                                "-an",
                                "`"$outputName`""
                            )
                            Description = "$fileNameWithoutExt | ${b}M | $preset | $tune | $($split.Name)"
                        }
                    }
                }
            }
        }
    }
}

$TotalJobs = $JobQueue.Count
Write-Host "Total Jobs Queued: $TotalJobs" -ForegroundColor Green
Start-Sleep -Seconds 2

# --- STEP 2: EXECUTE PARALLEL JOBS ---

# Initialize as an explicit array
$ActiveProcesses = @() 
$CompletedCount  = 0
$NextJobIndex    = 0

while ($CompletedCount -lt $TotalJobs) {
    
    # --- FIX IS HERE ---
    # We wrap the result in @(...) to force it to remain an array 
    # even if filtering leaves 0 or 1 items.
    $ActiveProcesses = @($ActiveProcesses | Where-Object { 
        if ($_.HasExited) {
            $global:CompletedCount++
            return $false # Remove from list
        }
        return $true # Keep in list
    })

    # Launch new jobs if slots are available
    while ($ActiveProcesses.Count -lt $MaxConcurrentJobs -and $NextJobIndex -lt $TotalJobs) {
        $job = $JobQueue[$NextJobIndex]
        
        $p = Start-Process -FilePath "ffmpeg" -ArgumentList $job.Args -NoNewWindow -PassThru
        
        # Now this addition works because $ActiveProcesses is guaranteed to be an array
        $ActiveProcesses += $p
        $NextJobIndex++
    }

    $percent = [math]::Round(($CompletedCount / $TotalJobs) * 100, 1)
    $running = $ActiveProcesses.Count
    
    Write-Progress -Activity "Blackwell Encoding Matrix" `
                   -Status "$percent% Complete ($CompletedCount / $TotalJobs)" `
                   -CurrentOperation "Active Encoders: $running | Queue Index: $NextJobIndex" `
                   -PercentComplete $percent

    Start-Sleep -Milliseconds 250
}

Write-Host "`nAll $TotalJobs encodes completed successfully." -ForegroundColor Green
\end{lstlisting}
\newpage
\section{Full Rate-Distortion Results}

Note: Sequences 22 and 23 are referred to as ``Basketball" in the official Netflix Chimera documentation, though the source media files and title screens are misspelled as ``Baskeyball". We use the corrected spelling in this text, but automated parsing scripts must account for the original file nomenclature.

\subsection{Peak Signal-to-Noise Ratio (PSNR)}

\begin{table}[!h]
\vspace{-2mm}
\setstretch{0.8}
\vspace{-0.5mm}
\caption{BD-Rate PSNR-Y (\%) (Pascal HQ Baseline)}
\vspace{0.5mm}
\centering
\resizebox{17.5cm}{!}{\begin{tabular}{@{}lcccccccccccccccccc@{}}
\toprule
\multirow{2.5}{*}{Sequence Name}& \multicolumn{3}{c}{Ampere HQ} & \multicolumn{3}{c}{Ampere UHQ} & \multicolumn{3}{c}{Ada Lovelace HQ} & \multicolumn{3}{c}{Ada Lovelace UHQ}& \multicolumn{3}{c}{Blackwell HQ}&\multicolumn{3}{c}{Blackwell UHQ}\\
\cmidrule(lr){2-4} \cmidrule(lr){5-7} \cmidrule(lr){8-10} \cmidrule(lr){11-13} \cmidrule(lr){14-16} \cmidrule(lr){17-19}
& P1 & P4 & P7 & P1 & P4 & P7 & P1 & P4 & P7 & P1 & P4 & P7 & P1 & P4 & P7 & P1 & P4 & P7\\\midrule
\multicolumn{19}{c}{\textbf{Netflix Chimera}}\\\midrule
1\_BAR\_SCENE&-13.98&-10.53&-10.28&-28.42&-24.86&-26.06&-13.57&-10.30&-9.54&-28.61&-24.57&-25.72&-17.43&-17.23&-16.78&-33.11&-35.26&-34.91                \\
2\_DINNER\_SCENE&-10.54&-5.30&-6.01&-37.30&-39.10&-43.27&-9.78&-5.24&-5.13&-37.12&-38.47&-42.59&-12.52&17.94&17.19&-41.27&-43.08&-45.55                   \\
3\_DANCER&22.54&21.37&22.59&-0.46&-2.86&-0.15&24.32&23.33&24.56&2.45&-0.21&2.48&21.58&19.03&19.68&-1.93&-7.58&-4.93                                       \\
4\_DANCERS\_COUPLE&3.88&5.76&7.84&-2.63&-3.18&-0.31&4.71&7.09&8.87&-0.91&-2.28&0.76&2.97&4.20&5.75&-6.48&-9.04&-6.56                                      \\
5\_DANCERS\_MONTAGE\_MIXED&-7.35&8.45&8.62&0.51&0.76&-1.36&-7.41&8.91&-3.91&1.11&2.73&-0.73&-3.22&21.83&5.90&3.79&9.23&8.06                               \\
6\_ROLLERCOASTER\_SEQUENCE&-6.85&-2.66&-0.84&-10.69&-5.00&-3.22&-7.06&-2.55&-0.76&-10.65&-4.87&-3.00&-8.35&-11.24&-9.84&-12.05&-15.36&-13.57              \\
7\_ROLLERCOASTER\_POV&-8.30&-3.95&-2.61&-11.92&-6.35&-4.82&-8.41&-3.84&-2.48&-11.82&-6.19&-4.71&-9.22&-11.59&-10.21&-12.62&-15.78&-14.27                  \\
8\_ROLLERCOASTER\_PASSENGER&-10.04&-5.43&-3.74&-14.08&-8.41&-7.07&-9.76&-5.41&-3.68&-14.11&-8.22&-6.79&-10.90&-14.14&-12.42&-15.71&-18.81&-17.22          \\
9\_TWIRL\_RIDE\_BOARDWALK&-10.59&-6.56&-5.76&-10.89&-5.46&-4.88&-10.41&-6.45&-5.68&-10.95&-5.46&-4.83&-14.43&-16.57&-15.50&-15.48&-16.85&-15.78           \\
10\_NETFLIX\_CARD\_TWIRL&-18.02&-17.88&-15.66&-26.40&-28.15&-27.36&-18.13&-17.90&-15.65&-26.42&-28.19&-27.44&-20.94&-22.68&-20.73&-30.41&-33.83&-33.10    \\
11\_SEASIDE\_AND\_PIER&-21.43&-17.04&-17.44&-28.44&-23.32&-23.84&-21.01&-16.52&-17.01&-28.05&-22.90&-23.45&-22.74&-23.39&-23.74&-30.48&-32.03&-32.40      \\
12\_WIND\_AND\_NATURE&-18.81&-16.55&-16.50&-26.71&-27.22&-30.27&-18.61&-16.43&-16.26&-26.62&-27.10&-30.14&-22.69&-22.73&-22.76&-30.83&-33.33&-35.96       \\
13\_MOUNTAIN\_VIEW\_W\_TILT&-28.35&-18.84&-19.46&-63.42&-55.81&-51.72&-26.62&-17.40&-17.75&-63.31&-55.14&-50.50&-23.91&-37.79&-37.96&-62.33&-69.36&-69.16 \\
14\_MOUNTAIN\_VIEW\_PAN&-27.02&-21.91&-22.97&-58.95&-50.14&-55.55&-26.36&-21.47&-22.86&-58.69&-49.16&-48.18&-25.23&-28.56&-29.72&-57.55&-61.03&-61.79     \\
15\_WALK\_LIKE\_A\_MAN&-24.00&-19.77&-20.11&-40.55&-38.85&-39.47&-23.59&-19.24&-19.68&-40.29&-38.59&-39.22&-22.64&-26.92&-26.89&-39.85&-43.98&-45.07      \\
16\_TODDLER\_AND\_FOUNTAIN&-4.73&-7.49&-6.53&-5.60&-8.50&-8.01&-4.70&-7.42&-6.47&-5.53&-8.41&-7.89&-11.34&-13.51&-13.20&-13.53&-15.76&-15.32              \\
17\_DRIVING\_POV&-12.05&-7.18&-7.15&-14.52&-8.90&-9.88&-11.46&-6.63&-6.59&-14.17&-8.55&-9.51&-13.61&-14.74&-14.58&-16.84&-18.26&-18.54                    \\
18\_PLANET\_MOBILE&-13.81&4.73&5.28&-23.75&-9.32&-10.95&-13.43&5.52&5.78&-23.69&-8.94&-10.57&-12.41&20.40&20.78&-22.94&-10.61&-12.63                      \\
19\_DOG\_PANTS&-15.98&-7.19&-8.45&-32.58&-25.76&-26.48&-15.95&-6.73&-7.48&-32.78&-25.78&-26.79&-14.52&-16.42&-17.41&-29.99&-31.68&-33.52                  \\
20\_DOG\_BARKS&-20.68&-16.52&-15.63&-39.38&-38.35&-40.07&-19.32&-14.81&-16.07&-39.47&-38.27&-40.12&-20.66&-21.29&-21.21&-41.53&-42.41&-43.07              \\
21\_RC\_AERIAL&-16.91&-12.09&-12.45&-27.84&-23.46&-24.20&-15.94&-11.29&-11.82&-27.23&-23.04&-23.71&-14.73&-16.25&-16.58&-24.94&-26.82&-27.92              \\
22\_BASKETBALL\_FREE\_THROW&-6.63&1.78&2.74&-11.28&-3.55&-3.43&-6.82&2.31&2.76&-11.21&-3.44&-3.08&-9.44&-5.76&-4.04&-15.83&-13.26&-12.83                  \\
23\_BASKETBALL\_GAME&-8.74&-7.13&-5.99&-6.24&-3.58&-2.86&-8.51&-7.09&-5.96&-6.35&-3.54&-2.84&-11.65&-14.69&-13.72&-10.60&-12.65&-11.08                    \\\midrule
\textbf{Average}&-12.10&-7.04&-6.54&-22.68&-19.10&-19.36&-11.64&-6.50&-6.64&-22.37&-18.63&-18.63&-12.96&-10.96&-11.22&-24.46&-25.98&-25.96                                                             \\\midrule
\multicolumn{19}{c}{\textbf{Twitch}}\\\midrule
CSGO&4.24&6.35&7.36&-0.77&0.70&0.75&4.47&6.80&7.67&-0.70&0.82&0.99&1.30&0.06&1.28&-5.63&-7.34&-6.73                                                       \\
DOTA2&-16.33&-13.16&-13.05&-27.50&-22.77&-23.64&-16.41&-13.20&-13.07&-27.57&-22.70&-23.52&-19.62&-20.54&-21.43&-29.61&-30.97&-31.36                       \\
EuroTruckSimulator2&-11.29&-7.51&-7.00&-16.14&-12.82&-13.56&-10.71&-7.12&-6.69&-15.94&-12.36&-13.24&-13.66&-13.06&-12.73&-19.33&-21.34&-21.89             \\
Fallout4&-5.94&-4.11&-3.61&-8.14&-6.10&-6.05&-5.67&-3.88&-3.38&-7.98&-5.92&-5.80&-9.46&-10.51&-10.15&-11.64&-13.37&-13.11                                 \\
GTAV&-4.04&1.96&2.81&-5.44&0.23&-0.23&-3.71&2.21&3.06&-5.36&0.34&-0.04&-7.57&-5.28&-4.96&-10.13&-9.86&-9.71                                               \\
Hearthstone&-12.27&-11.72&-10.81&-47.29&-47.51&-50.32&-12.45&-9.94&-11.14&-47.33&-47.58&-50.13&-13.54&-14.60&-14.17&-47.18&-49.56&-52.46                  \\
MINECRAFT&9.52&16.64&20.28&7.42&9.30&11.31&9.12&16.31&20.03&7.04&9.33&11.22&7.79&10.45&13.49&3.08&0.70&3.79                                               \\
RUST&-7.61&-1.37&-0.88&-5.97&-0.81&-1.83&-7.43&-1.20&-0.69&-5.98&-0.69&-1.69&-11.55&-10.12&-9.70&-11.65&-12.09&-12.78                                     \\
STARCRAFT&-13.28&-8.68&-7.68&-31.76&-27.93&-30.97&-13.48&-8.54&-7.78&-31.73&-27.86&-31.00&-14.21&-11.31&-10.88&-33.65&-33.08&-35.66                       \\
WITCHER3&-9.66&-3.04&-2.89&-13.10&-6.35&-7.32&-9.69&-2.93&-3.28&-12.79&-6.68&-7.25&-12.58&-9.67&-10.01&-18.33&-16.58&-16.27                               \\\midrule
\textbf{Average}&-6.67&-2.46&-1.55&-14.87&-11.41&-12.19&-6.60&-2.15&-1.53&-14.83&-11.33&-12.05&-9.31&-8.46&-7.93&-18.41&-19.35&-19.62                              \\
\bottomrule
\end{tabular}}
\vspace{-4mm}
\end{table}

\begin{table}[!h]
\vspace{-2mm}
\setstretch{0.8}
\vspace{-0.5mm}
\caption{BD-Rate PSNR-Y (\%) (Ada Lovelace HQ Baseline)}
\vspace{0.5mm}
\centering
\resizebox{17.5cm}{!}{\begin{tabular}{@{}lcccccccccccccccccc@{}}
\toprule
\multirow{2.5}{*}{Sequence Name}& \multicolumn{3}{c}{Ampere HEVC HQ} & \multicolumn{3}{c}{Ampere HEVC UHQ} & \multicolumn{3}{c}{Blackwell HEVC HQ} & \multicolumn{3}{c}{Blackwell HEVC UHQ}& \multicolumn{3}{c}{Blackwell AV1 HQ}&\multicolumn{3}{c}{Blackwell AV1 UHQ}\\
\cmidrule(lr){2-4} \cmidrule(lr){5-7} \cmidrule(lr){8-10} \cmidrule(lr){11-13} \cmidrule(lr){14-16} \cmidrule(lr){17-19}
& P1 & P4 & P7 & P1 & P4 & P7 & P1 & P4 & P7 & P1 & P4 & P7 & P1 & P4 & P7 & P1 & P4 & P7\\\midrule
\multicolumn{19}{c}{\textbf{Netflix Chimera}}\\\midrule
1\_BAR\_SCENE&-0.51&-0.28&-0.84&-16.54&-15.45&-17.60&-4.22&-6.94&-7.26&-17.18&-21.16&-22.17&1.24&3.24&2.77&-23.33&-22.05&-22.01                \\
2\_DINNER\_SCENE&-0.95&-0.20&-1.01&-32.28&-36.77&-41.46&-4.38&22.51&21.40&-35.22&-38.31&-41.70&21.24&23.06&22.51&-21.83&-21.21&-21.47          \\
3\_DANCER&-1.50&-1.52&-1.53&-18.77&-17.73&-16.38&-2.16&-3.08&-3.55&-11.28&-14.96&-14.76&3.91&4.59&4.58&8.10&3.27&-0.06                         \\
4\_DANCERS\_COUPLE&-0.80&-1.25&-0.97&-6.72&-9.13&-8.09&-2.18&-2.80&-3.00&-9.47&-11.77&-11.35&1.14&2.13&2.08&-2.42&-3.95&-5.91                  \\
5\_DANCERS\_MONTAGE\_MIXED&0.00&-0.44&11.06&8.81&-7.67&3.15&4.55&11.44&9.78&22.60&10.72&25.07&11.24&13.62&13.76&9.25&8.90&7.17                 \\
6\_ROLLERCOASTER\_SEQUENCE&0.21&-0.12&-0.07&-3.58&-2.23&-2.17&-1.46&-9.04&-9.26&-7.90&-13.03&-13.13&-2.78&-1.68&-1.64&-14.68&-14.56&-14.97     \\
7\_ROLLERCOASTER\_POV&0.10&-0.12&-0.14&-3.48&-2.35&-2.15&-0.97&-8.21&-8.12&-6.71&-11.86&-11.99&-2.22&-1.11&-0.94&-16.49&-16.55&-16.80          \\
8\_ROLLERCOASTER\_PASSENGER&-0.31&-0.04&-0.07&-4.34&-3.01&-3.38&-1.34&-9.25&-9.11&-5.39&-11.55&-11.84&-2.68&-1.65&-1.67&-14.89&-15.35&-15.80   \\
9\_TWIRL\_RIDE\_BOARDWALK&-0.20&-0.12&-0.08&-0.33&0.94&0.73&-4.52&-10.84&-10.47&-2.70&-6.67&-6.30&-4.31&-2.81&-2.49&-15.78&-15.44&-14.84       \\
10\_NETFLIX\_CARD\_TWIRL&0.12&0.01&-0.03&-9.96&-12.06&-13.51&-3.42&-5.78&-5.99&-12.79&-17.35&-18.74&-2.04&-1.64&-1.55&-22.70&-23.82&-23.95     \\
11\_SEASIDE\_AND\_PIER&-0.54&-0.61&-0.51&-9.04&-7.78&-7.91&-2.48&-8.43&-8.34&-6.27&-11.16&-11.44&-3.01&-1.90&-2.04&-19.62&-19.01&-18.75        \\
12\_WIND\_AND\_NATURE&-0.24&-0.14&-0.28&-9.66&-12.76&-16.52&-5.07&-7.56&-7.79&-24.69&-28.13&-30.86&-1.16&-0.99&-0.84&-27.30&-29.37&-29.66      \\
13\_MOUNTAIN\_VIEW\_W\_TILT&-2.33&-1.70&-2.02&-48.58&-45.63&-40.49&3.63&-24.27&-24.03&-47.52&-58.89&-59.03&2.76&3.86&3.62&-49.86&-49.62&-48.49 \\
14\_MOUNTAIN\_VIEW\_PAN&-0.89&-0.52&-0.14&-42.21&-35.36&-41.17&1.43&-8.36&-8.08&-37.47&-42.93&-43.34&10.93&11.52&11.75&-46.46&-46.56&-47.12    \\
15\_WALK\_LIKE\_A\_MAN&-0.54&-0.64&-0.53&-21.86&-23.75&-24.12&1.00&-9.29&-8.79&-15.38&-22.29&-23.62&0.69&1.29&1.29&-28.90&-29.91&-30.40        \\
16\_TODDLER\_AND\_FOUNTAIN&-0.03&-0.08&-0.05&-0.92&-1.19&-1.63&-6.93&-6.41&-7.02&-15.54&-17.20&-17.65&-3.17&-2.86&-2.80&-10.89&-10.90&-10.18   \\
17\_DRIVING\_POV&-0.66&-0.58&-0.59&-3.26&-2.52&-3.63&-2.48&-8.74&-8.60&-2.91&-7.90&-8.39&-3.42&-2.36&-2.17&-17.42&-17.09&-16.91                \\
18\_PLANET\_MOBILE&-0.42&-0.71&-0.44&-12.01&-13.72&-15.36&1.53&13.17&13.34&-14.12&-17.34&-19.29&-0.78&0.84&1.00&-3.52&-4.36&-6.10              \\
19\_DOG\_PANTS&-0.05&-0.45&-0.97&-19.47&-20.41&-20.68&1.53&-10.36&-10.66&-14.28&-21.09&-22.83&2.25&3.88&3.68&-24.57&-24.56&-25.52              \\
20\_DOG\_BARKS&-1.70&-1.98&0.40&-24.23&-26.78&-27.80&-1.85&-7.55&-6.26&-25.68&-29.79&-30.46&1.80&2.74&2.66&-32.16&-32.61&-32.73                \\
21\_RC\_AERIAL&-1.13&-0.88&-0.71&-13.77&-13.49&-13.80&1.26&-5.65&-5.46&-5.88&-10.54&-11.47&-1.70&-0.88&-0.79&-20.28&-20.63&-20.93              \\
22\_BASKETBALL\_FREE\_THROW&0.18&-0.45&0.02&-4.71&-5.45&-5.70&-2.81&-7.85&-6.70&-12.57&-16.17&-16.70&-2.55&-1.64&-1.46&-13.34&-12.88&-14.27    \\
23\_BASKETBALL\_GAME&-0.25&0.00&0.01&2.71&3.87&3.39&-3.45&-8.13&-8.21&0.60&-1.95&-1.90&-3.81&-2.82&-2.83&-11.79&-11.12&-11.22                  \\\midrule
\textbf{Average}&-0.54&-0.56&0.02&-12.79&-13.50&-13.75&-1.51&-5.28&-5.31&-13.38&-18.32&-18.43&1.02&2.11&2.11&-18.30&-18.67&-19.17                       \\\midrule
\multicolumn{19}{c}{\textbf{Twitch}}\\\midrule
CSGO&-0.21&-0.42&-0.29&-2.32&-0.93&-1.20&-2.94&-6.30&-5.94&-5.65&-8.23&-8.26&-2.11&-1.04&-0.89&-18.80&-18.21&-18.42                            \\
DOTA2&0.07&0.03&0.01&-11.56&-9.35&-11.27&-4.04&-8.46&-9.60&-13.71&-17.30&-18.37&-3.13&-2.79&-2.90&-26.68&-27.12&-27.58                         \\
EuroTruckSimulator2&-0.64&-0.42&-0.34&-4.51&-9.70&-11.29&-3.36&-6.59&-6.66&-8.88&-15.05&-16.12&-2.09&-1.62&-1.45&-23.94&-24.96&-22.03          \\
Fallout4&-0.28&-0.23&-0.23&-6.13&-4.87&-5.50&-4.07&-7.08&-7.20&-8.90&-11.55&-11.83&-3.03&-2.26&-2.04&-16.95&-16.49&-16.89                      \\
GTAV&-0.35&-0.25&-0.24&-1.90&-3.02&-4.77&-3.97&-7.50&-7.97&-7.49&-12.92&-13.84&-3.24&-1.74&-1.90&-11.24&-19.87&-11.43                          \\
Hearthstone&0.19&-1.87&0.34&-28.27&-30.22&-33.08&-1.21&-5.21&-3.46&-29.56&-34.16&-37.19&2.22&-4.10&2.62&-34.21&-40.45&-35.49                   \\
MINECRAFT&0.35&0.27&0.20&-5.38&-7.06&-8.65&-1.35&-5.20&-5.60&3.81&-4.35&-4.39&0.07&0.81&1.54&-14.13&-8.58&-15.60                               \\
RUST&-0.20&-0.16&-0.18&1.82&0.47&-1.06&-4.47&-8.94&-8.98&-4.30&-10.53&-11.73&-4.15&-3.36&-3.40&-13.27&-13.87&-14.46                            \\
STARCRAFT&0.21&-0.17&0.12&-11.56&-12.25&-16.80&-1.04&-3.20&-3.52&-16.16&-20.71&-24.34&-1.48&-1.17&-1.05&-26.36&-27.12&-27.45                   \\
WITCHER3&0.03&-0.11&0.39&-2.22&-2.47&-3.58&-3.21&-6.67&-6.76&-7.39&-11.63&-11.48&-3.44&-2.37&-2.47&-17.46&-17.32&-17.21                        \\\midrule
\textbf{Average}&-0.08&-0.33&-0.02&-7.20&-7.94&-9.72&-2.97&-6.52&-6.57&-9.82&-14.64&-15.76&-2.04&-1.96&-1.19&-20.30&-21.40&-20.66                       \\
\bottomrule
\end{tabular}}
\vspace{-4mm}
\end{table}

\begin{table}[!h]
\vspace{-2mm}
\setstretch{0.8}
\vspace{-0.5mm}
\caption{BD-Rate PSNR-Y (\%) (HQ vs. UHQ)}
\vspace{0.5mm}
\centering
\resizebox{17.5cm}{!}{\begin{tabular}{@{}lcccccccccccccccccc@{}}
\toprule
\multirow{2.5}{*}{Sequence Name}& \multicolumn{3}{c}{Ampere HEVC} & \multicolumn{3}{c}{Ada Lovelace HEVC} & \multicolumn{3}{c}{Blackwell HEVC} & \multicolumn{3}{c}{Ada Lovelace AV1}& \multicolumn{3}{c}{Blackwell AV1}\\
\cmidrule(lr){2-4} \cmidrule(lr){5-7} \cmidrule(lr){8-10} \cmidrule(lr){11-13} \cmidrule(lr){14-16} 
& P1 & P4 & P7 & P1 & P4 & P7 & P1 & P4 & P7 & P1 & P4 & P7 & P1 & P4 & P7\\\midrule
\multicolumn{16}{c}{\textbf{Netflix Chimera}}\\\midrule
1\_BAR\_SCENE&-16.11&-15.19&-16.84&-16.74&-15.12&-17.17&-13.35&-15.15&-15.92&-20.80&-19.61&-19.78&-23.31&-23.01&-23.02             \\
2\_DINNER\_SCENE&-31.54&-36.60&-40.68&-31.89&-35.89&-40.58&-33.45&-44.19&-46.66&-25.43&-26.92&-27.50&-29.06&-29.03&-29.73          \\
3\_DANCER&-17.90&-16.74&-15.32&-17.10&-16.07&-14.67&-10.04&-12.35&-11.81&-0.95&-3.71&-4.27&4.03&-1.33&-4.52                       \\
4\_DANCERS\_COUPLE&-5.97&-8.00&-7.21&-5.01&-8.30&-7.10&-7.44&-9.23&-8.66&-5.00&-6.38&-6.66&-3.01&-5.42&-7.44                       \\
5\_DANCERS\_MONTAGE\_MIXED&8.76&-7.30&-8.54&9.44&-6.01&3.76&18.35&-1.08&13.64&-8.64&-9.38&-10.29&-0.65&-2.77&-4.53                  \\
6\_ROLLERCOASTER\_SEQUENCE&-3.79&-2.11&-2.09&-3.52&-2.10&-1.95&-6.52&-4.43&-4.26&-9.34&-10.13&-10.21&-12.30&-13.04&-13.48          \\
7\_ROLLERCOASTER\_POV&-3.58&-2.24&-2.01&-3.35&-2.20&-2.03&-5.75&-4.05&-4.25&-10.91&-11.78&-11.82&-14.64&-15.48&-15.91              \\
8\_ROLLERCOASTER\_PASSENGER&-4.03&-2.98&-3.30&-4.35&-2.82&-3.09&-4.08&-2.53&-3.01&-9.29&-10.42&-10.70&-12.40&-13.68&-14.16         \\
9\_TWIRL\_RIDE\_BOARDWALK&-0.12&1.06&0.81&-0.37&0.95&0.78&1.89&4.08&3.93&-9.78&-10.67&-9.88&-11.84&-12.70&-12.42                    \\
10\_NETFLIX\_CARD\_TWIRL&-10.05&-12.04&-13.46&-9.96&-12.08&-13.57&-9.75&-12.30&-13.58&-17.20&-19.04&-19.24&-21.24&-22.63&-22.86     \\
11\_SEASIDE\_AND\_PIER&-8.54&-7.21&-7.43&-8.55&-7.28&-7.44&-3.85&-3.05&-3.42&-15.17&-15.14&-15.08&-16.49&-16.71&-16.51              \\
12\_WIND\_AND\_NATURE&-9.42&-12.63&-16.26&-9.54&-12.62&-16.36&-20.52&-22.04&-24.82&-17.90&-20.85&-21.21&-26.13&-28.32&-28.75        \\
13\_MOUNTAIN\_VIEW\_W\_TILT&-47.09&-44.64&-39.22&-48.44&-44.84&-39.19&-49.49&-46.70&-46.97&-46.00&-47.61&-46.79&-52.95&-53.27&-51.96 \\
14\_MOUNTAIN\_VIEW\_PAN&-41.73&-34.93&-40.97&-41.84&-34.02&-31.75&-37.90&-37.03&-37.33&-38.70&-38.87&-38.39&-51.93&-52.27&-52.71    \\
15\_WALK\_LIKE\_A\_MAN&-21.41&-23.26&-23.70&-21.52&-23.41&-23.78&-16.21&-14.43&-16.31&-25.86&-26.78&-27.82&-30.05&-31.49&-31.90      \\
16\_TODDLER\_AND\_FOUNTAIN&-0.89&-1.10&-1.58&-0.83&-1.09&-1.47&-9.03&-11.33&-11.09&-1.55&-2.22&-2.48&-7.83&-8.11&-7.45              \\
17\_DRIVING\_POV&-2.61&-1.95&-3.06&-2.86&-2.13&-3.21&-0.52&0.72&-0.02&-12.62&-12.82&-12.59&-14.68&-15.22&-15.23                    \\
18\_PLANET\_MOBILE&-11.62&-13.05&-14.94&-12.00&-13.42&-15.08&-15.52&-25.71&-27.44&-5.12&-5.30&-5.96&-3.55&-5.67&-7.73              \\
19\_DOG\_PANTS&-19.43&-20.02&-19.86&-19.60&-20.49&-21.03&-15.57&-12.01&-13.74&-23.34&-23.77&-24.05&-26.50&-27.34&-28.26            \\
20\_DOG\_BARKS&-22.70&-25.34&-27.97&-24.30&-26.63&-27.80&-24.19&-23.99&-25.71&-23.57&-24.48&-25.75&-34.04&-35.23&-35.24            \\
21\_RC\_AERIAL&-12.77&-12.70&-13.16&-13.03&-12.97&-13.21&-7.10&-5.24&-6.41&-16.48&-18.30&-18.65&-19.29&-20.18&-20.60               \\
22\_BASKETBALL\_FREE\_THROW&-4.85&-4.99&-5.70&-4.63&-5.37&-5.42&-10.04&-9.03&-10.77&-8.60&-8.47&-9.18&-11.25&-11.69&-13.13          \\
23\_BASKETBALL\_GAME&2.97&3.88&3.38&2.63&3.89&3.36&4.21&6.59&6.74&-7.04&-7.23&-7.01&-8.25&-8.43&-8.51                              \\\midrule
\textbf{Average}&-12.37&-13.05&-13.87&-12.49&-13.04&-12.96&-11.99&-13.24&-13.39&-15.62&-16.52&-16.75&-18.58&-19.70&-20.26                 \\\midrule
\multicolumn{16}{c}{\textbf{Twitch}}\\\midrule
CSGO&-4.90&-5.27&-6.11&-5.06&-5.56&-6.16&-6.98&-7.42&-7.93&-11.97&-12.15&-12.42&-13.38&-13.60&-14.02                             \\
DOTA2&-12.92&-10.54&-11.63&-12.92&-10.40&-11.46&-12.10&-13.02&-12.60&-22.65&-23.09&-23.32&-23.50&-24.05&-24.68                   \\
EuroTruckSimulator2&-5.26&-5.75&-7.13&-5.61&-5.64&-7.08&-6.63&-9.42&-10.41&-16.06&-16.92&-16.85&-18.04&-19.25&-19.70             \\
Fallout4&-2.31&-1.95&-2.40&-2.43&-1.99&-2.38&-2.24&-2.83&-2.91&-8.16&-8.74&-9.08&-8.65&-9.01&-9.52                               \\
GTAV&-1.53&-1.65&-2.92&-1.79&-1.78&-2.97&-2.82&-4.42&-4.47&-7.14&-7.87&-8.40&-8.42&-10.00&-9.77                                  \\
Hearthstone&-39.04&-39.87&-43.61&-39.00&-41.08&-43.28&-38.07&-40.10&-43.91&-23.88&-30.54&-25.45&-28.83&-29.92&-30.91             \\
MINECRAFT&-1.96&-6.05&-7.14&-1.97&-5.83&-7.05&-4.56&-8.55&-8.21&-7.01&-8.61&-10.22&-7.45&-8.61&-9.72                             \\
RUST&1.83&0.56&-0.90&1.63&0.51&-0.94&-0.09&-2.16&-3.33&-7.33&-8.73&-9.37&-8.54&-9.86&-10.58                                      \\
STARCRAFT&-21.38&-21.23&-25.46&-21.12&-21.30&-25.44&-22.59&-24.71&-27.96&-20.57&-21.21&-22.73&-24.52&-25.15&-26.05               \\
WITCHER3&-3.73&-3.37&-4.52&-3.36&-3.80&-4.06&-6.54&-7.66&-7.01&-11.32&-11.36&-11.63&-13.71&-14.21&-14.05                         \\\midrule
\textbf{Average}&-9.12&-9.51&-11.18&-9.16&-9.69&-11.08&-10.26&-12.03&-12.87&-13.61&-14.92&-14.95&-15.50&-16.37&-16.90                     \\
\bottomrule
\end{tabular}}
\vspace{-4mm}
\end{table}

\begin{table}[!h]
\vspace{-2mm}
\setstretch{0.8}
\vspace{-0.5mm}
\caption{BD-Rate PSNR-Y (\%) (Ada Lovelace UHQ Baseline)}
\vspace{0.5mm}
\centering
\resizebox{12cm}{!}{\begin{tabular}{@{}lcccccccccccccccccc@{}}
\toprule
\multirow{2.5}{*}{Sequence Name}& \multicolumn{3}{c}{Ampere HEVC UHQ} & \multicolumn{3}{c}{Blackwell HEVC UHQ} & \multicolumn{3}{c}{Blackwell AV1 UHQ} \\
\cmidrule(lr){2-4} \cmidrule(lr){5-7} \cmidrule(lr){8-10} 
& P1 & P4 & P7 & P1 & P4 & P7 & P1 & P4 & P7 \\\midrule
\multicolumn{10}{c}{\textbf{Netflix Chimera}}\\\midrule
1\_BAR\_SCENE&0.43&-0.02&-0.08&-11.12&-17.87&-14.41&3.07&1.28&2.00                  \\
2\_DINNER\_SCENE&-0.19&-0.67&-0.91&-14.91&-13.70&-9.46&13.67&16.43&17.37            \\
3\_DANCER&-1.62&-1.72&-1.64&-10.28&-10.47&-9.79&10.56&7.57&4.76                     \\
4\_DANCERS\_COUPLE&-1.26&-0.55&-0.73&-7.61&-8.08&-8.01&9.62&9.54&7.66               \\
5\_DANCERS\_MONTAGE\_MIXED&-0.33&-1.21&-0.31&-5.11&-1.38&0.57&17.94&19.18&17.89     \\
6\_ROLLERCOASTER\_SEQUENCE&0.25&0.01&-0.10&0.64&-8.83&-7.82&-2.40&-1.53&-2.24       \\
7\_ROLLERCOASTER\_POV&0.02&-0.03&0.05&0.66&-7.95&-7.18&-2.60&-1.77&-2.66            \\
8\_ROLLERCOASTER\_PASSENGER&0.23&-0.02&-0.11&-3.15&-12.69&-11.71&-3.10&-2.20&-2.68  \\
9\_TWIRL\_RIDE\_BOARDWALK&0.28&0.21&0.09&-7.56&-14.07&-13.41&-4.92&-3.28&-3.63      \\
10\_NETFLIX\_CARD\_TWIRL&0.06&0.12&0.23&-8.88&-11.15&-10.98&-3.90&-3.60&-3.66       \\
11\_SEASIDE\_AND\_PIER&-0.34&-0.30&-0.25&-7.72&-15.15&-14.60&-2.64&-2.06&-2.02      \\
12\_WIND\_AND\_NATURE&0.02&-0.03&-0.12&5.98&2.09&2.69&2.81&3.43&3.77                \\
13\_MOUNTAIN\_VIEW\_W\_TILT&0.68&-1.41&-1.81&2.56&-22.88&-28.18&7.45&17.10&15.63    \\
14\_MOUNTAIN\_VIEW\_PAN&-0.21&0.24&-21.81&-7.34&-31.64&-32.20&7.43&6.80&6.72        \\
15\_WALK\_LIKE\_A\_MAN&-0.24&-0.23&-0.15&-5.09&-13.37&-13.86&-0.92&-0.58&-0.90      \\
16\_TODDLER\_AND\_FOUNTAIN&0.04&-0.04&-0.06&-6.56&-1.83&-2.19&-3.91&-2.59&-3.00     \\
17\_DRIVING\_POV&-0.22&-0.22&-0.26&-5.50&-11.95&-11.16&-3.73&-2.77&-2.96            \\
18\_PLANET\_MOBILE&-0.02&-0.33&-0.39&1.22&-0.35&-0.57&5.93&5.67&4.67                \\
19\_DOG\_PANTS&0.44&0.30&0.62&-2.18&-9.53&-10.14&2.93&3.38&2.39                     \\
20\_DOG\_BARKS&0.28&0.29&-0.17&-3.69&-8.65&-7.12&2.93&3.70&4.33                     \\
21\_RC\_AERIAL&-0.41&-0.11&-0.20&-2.45&-10.30&-10.80&3.47&-4.54&-4.79               \\
22\_BASKETBALL\_FREE\_THROW&0.11&-0.08&0.23&-2.14&-5.96&-5.44&2.54&3.51&3.52        \\
23\_BASKETBALL\_GAME&0.43&-4.83&-0.07&-1.38&-10.36&-9.41&-4.32&-3.31&-3.78          \\\midrule
\textbf{Average}&-0.07&-0.46&-1.22&-4.42&-10.70&-10.23&2.52&3.02&2.54                        \\\midrule
\multicolumn{10}{c}{\textbf{Twitch}}\\\midrule
CSGO&3.00&5.00&5.26&-4.99&-8.11&-7.69&-3.80&-2.91&-2.85                             \\
DOTA2&1.55&1.17&0.22&-3.35&-10.89&-10.50&-4.15&-4.01&-4.32                          \\
EuroTruckSimulator2&1.23&-4.12&-4.32&-4.64&-10.35&-10.02&-4.18&-4.13&-4.43          \\
Fallout4&-3.53&-2.74&-3.00&-4.03&-7.95&-7.79&-3.48&-2.62&-2.53                      \\
GTAV&-0.02&-1.18&-1.78&-5.14&-10.22&-9.69&-4.68&-4.54&-3.78                         \\
Hearthstone&16.90&17.40&17.03&0.96&-2.33&-3.13&-4.12&-4.49&-4.59                    \\
MINECRAFT&-3.39&-1.03&-1.55&-3.71&-7.92&-6.75&-0.48&0.64&1.45                       \\
RUST&0.27&0.04&-0.05&-6.13&-11.39&-11.21&-5.39&-4.70&-4.70                          \\
STARCRAFT&11.79&11.12&11.11&-2.84&-6.99&-6.43&-5.66&-5.54&-4.82                     \\
WITCHER3&1.19&1.39&0.50&-6.38&-9.97&-9.06&-6.24&-6.07&-5.59                         \\\midrule
\textbf{Average}&2.90&2.71&2.34&-4.03&-8.61&-8.23&-4.22&-3.84&-3.62                          \\
\bottomrule
\end{tabular}}
\vspace{-4mm}
\end{table}

\clearpage

\subsection{Video Multimethod Assessment Fusion (VMAF) 4K - No Enhancement Gain (NEG)}

\begin{table}[!h]
\vspace{-2mm}
\setstretch{0.8}
\vspace{-0.5mm}
\caption{BD-Rate VMAF 4K NEG (\%) (Pascal HQ Baseline)}
\vspace{0.5mm}
\centering
\resizebox{17.5cm}{!}{\begin{tabular}{@{}lcccccccccccccccccc@{}}
\toprule
\multirow{2.5}{*}{Sequence Name}& \multicolumn{3}{c}{Ampere HQ} & \multicolumn{3}{c}{Ampere UHQ} & \multicolumn{3}{c}{Ada Lovelace HQ} & \multicolumn{3}{c}{Ada Lovelace UHQ}& \multicolumn{3}{c}{Blackwell HQ}&\multicolumn{3}{c}{Blackwell UHQ}\\
\cmidrule(lr){2-4} \cmidrule(lr){5-7} \cmidrule(lr){8-10} \cmidrule(lr){11-13} \cmidrule(lr){14-16} \cmidrule(lr){17-19}
& P1 & P4 & P7 & P1 & P4 & P7 & P1 & P4 & P7 & P1 & P4 & P7 & P1 & P4 & P7 & P1 & P4 & P7\\\midrule
\multicolumn{19}{c}{\textbf{Netflix Chimera}}\\\midrule
1\_BAR\_SCENE&-7.55&-3.57&-3.40&-20.43&-14.17&-15.59&-7.13&-3.46&-2.64&-20.52&-13.64&-14.95&-11.25&-10.40&-10.02&-24.61&-25.85&-26.36                \\
2\_DINNER\_SCENE&6.21&12.36&11.95&-21.90&-23.26&-29.32&7.81&13.12&13.52&-21.74&-21.96&-28.12&6.69&10.01&51.17&-32.89&-36.10&-39.68                   \\
3\_DANCER&78.78&90.30&88.91&45.30&50.76&49.38&85.63&95.37&95.04&53.09&55.99&54.44&72.55&75.76&74.81&16.09&18.96&23.53                               \\
4\_DANCERS\_COUPLE&21.35&30.59&31.60&18.33&19.63&21.57&22.70&32.22&32.95&20.89&20.72&23.17&19.06&26.77&27.87&5.88&5.06&7.63                          \\
5\_DANCERS\_MONTAGE\_MIXED&63.91&43.62&43.14&32.46&28.89&27.76&63.91&40.77&40.73&33.74&32.64&28.51&22.71&40.10&41.83&12.91&14.68&17.19                \\
6\_ROLLERCOASTER\_SEQUENCE&3.06&7.90&9.75&0.71&7.47&9.70&2.78&8.09&9.85&0.87&7.62&10.02&2.44&-2.59&-1.07&0.11&-5.14&-3.17                            \\
7\_ROLLERCOASTER\_POV&2.58&7.89&8.95&2.04&8.95&10.69&2.49&7.95&9.08&2.28&9.20&10.82&2.62&-0.91&0.21&1.89&-2.89&-1.25                                 \\
8\_ROLLERCOASTER\_PASSENGER&-1.60&4.04&5.37&-6.01&0.46&1.85&-1.21&4.08&5.53&-5.87&0.80&2.24&-1.28&-5.97&-4.40&-6.76&-11.16&-9.75                     \\
9\_TWIRL\_RIDE\_BOARDWALK&-1.41&3.81&4.15&-0.12&6.62&6.99&-1.13&3.97&4.22&-0.01&6.69&7.10&-4.08&-6.88&-6.16&-3.87&-5.61&-4.68                         \\
10\_NETFLIX\_CARD\_TWIRL&-12.77&-11.50&-8.87&-21.00&-20.43&-19.04&-12.84&-11.50&-8.89&-20.96&-20.41&-19.10&-15.75&-16.24&-13.88&-24.15&-25.45&-24.12  \\
11\_SEASIDE\_AND\_PIER&-6.46&-2.86&-3.10&-7.23&-2.92&-3.49&-5.73&-2.21&-2.55&-6.63&-2.28&-2.87&-9.69&-11.46&-11.77&-10.43&-13.26&-13.94               \\
12\_WIND\_AND\_NATURE&-13.86&-12.47&-12.79&-20.95&-21.79&-25.11&-13.71&-12.27&-12.60&-20.81&-21.62&-24.93&-17.95&-18.54&-18.96&-24.88&-27.00&-30.09   \\
13\_MOUNTAIN\_VIEW\_W\_TILT&-16.96&-7.12&-7.32&-50.98&-40.49&-36.44&-14.75&-5.23&-5.16&-50.26&-39.47&-34.98&-12.59&-28.77&-28.40&-50.23&-55.37&-56.51  \\
14\_MOUNTAIN\_VIEW\_PAN&-15.89&-9.17&-10.25&-45.48&-29.20&-39.69&-14.67&-8.91&-10.63&-44.93&-28.19&-27.73&-14.97&-17.03&-17.76&-45.17&-46.06&-47.64   \\
15\_WALK\_LIKE\_A\_MAN&-15.54&-11.81&-12.23&-28.55&-25.67&-26.99&-14.91&-11.02&-11.58&-28.01&-25.53&-26.63&-15.95&-20.54&-20.48&-28.60&-30.31&-31.94   \\
16\_TODDLER\_AND\_FOUNTAIN&-0.13&-2.56&-0.34&-0.62&-3.04&-1.56&0.15&-2.29&-0.08&-0.32&-2.74&-1.25&-8.71&-12.84&-11.24&-10.32&-14.49&-13.19            \\
17\_DRIVING\_POV&-4.78&1.61&1.50&-4.83&2.26&1.06&-4.13&2.20&2.11&-4.35&2.75&1.52&-6.14&-6.92&-6.80&-6.67&-7.73&-8.21                                 \\
18\_PLANET\_MOBILE&2.99&20.71&20.88&-16.54&-7.31&-11.58&3.75&22.11&21.85&-16.51&-6.46&-10.99&2.23&28.87&29.34&-28.56&-33.79&-35.57                   \\
19\_DOG\_PANTS&-0.82&8.96&7.74&-15.47&-7.48&-7.69&-0.23&9.65&9.42&-15.34&-7.14&-8.00&1.11&-3.95&-4.71&-13.39&-13.82&-16.04                           \\
20\_DOG\_BARKS&-13.13&-10.96&-8.77&-28.77&-29.32&-30.93&-11.31&-8.45&-9.06&-28.68&-29.02&-30.85&-14.19&-16.23&-15.24&-32.31&-34.17&-34.72            \\
21\_RC\_AERIAL&-4.48&1.76&1.59&-11.73&-5.35&-6.13&-3.14&2.65&2.26&-10.89&-4.84&-5.56&-3.68&-5.10&-5.38&-9.57&-10.09&-11.42                           \\
22\_BASKETBALL\_FREE\_THROW&5.57&13.46&14.52&0.97&6.65&7.00&5.68&14.45&14.63&1.34&7.01&7.60&2.98&2.88&5.05&-6.43&-7.65&-6.93                          \\
23\_BASKETBALL\_GAME&0.94&4.28&5.51&4.14&8.71&10.40&1.26&4.09&5.39&4.16&8.78&10.41&-0.83&-4.62&-3.38&0.90&-1.14&1.06                                 \\
\midrule
\textbf{Average}&3.04&7.79&8.20&-8.55&-3.91&-4.66&3.97&8.49&8.84&-7.80&-3.09&-3.48&-0.20&-0.20&2.20&-13.96&-16.02&-15.90                                    \\\midrule
\multicolumn{19}{c}{\textbf{Twitch}}\\\midrule
CSGO&19.07&20.76&22.10&16.44&17.98&18.07&19.57&21.46&22.62&16.77&18.28&18.46&15.26&12.10&13.89&11.26&7.54&8.55                                     \\
DOTA2&-10.11&-6.51&-6.20&-15.38&-7.88&-8.75&-10.15&-6.38&-5.96&-15.37&-7.82&-8.57&-14.42&-15.86&-16.52&-17.27&-19.75&-19.88                        \\
EuroTruckSimulator2&-3.71&0.21&0.38&-6.21&-3.25&-4.56&-2.84&0.91&1.00&-5.66&-2.51&-3.98&-7.92&-7.57&-7.58&-10.38&-13.05&-14.09                     \\
Fallout4&5.10&5.66&6.40&4.04&5.73&6.35&5.69&6.10&6.83&4.53&6.18&6.84&1.76&-1.29&-0.68&1.23&-2.35&-1.87                                             \\
GTAV&7.66&11.90&12.72&6.76&12.56&12.21&8.32&12.78&13.43&7.21&12.99&12.77&4.03&2.16&2.29&2.08&-0.73&-0.92                                           \\
Hearthstone&-2.04&0.47&2.79&-35.22&-33.92&-36.65&-2.54&2.89&2.60&-35.29&-34.04&-36.31&-2.91&-3.12&-1.79&-35.53&-35.43&-38.42                       \\
MINECRAFT&24.43&31.26&35.25&19.89&21.30&23.96&24.19&31.13&35.23&19.41&21.49&24.01&22.14&23.42&26.50&16.17&10.00&13.79                              \\
RUST&3.47&7.47&7.34&6.66&9.18&7.50&3.89&7.90&7.70&6.97&9.53&7.84&-2.99&-5.95&-6.23&-2.18&-7.77&-9.10                                               \\
STARCRAFT&-3.91&2.44&4.52&-19.62&-13.40&-17.50&-4.45&2.77&3.83&-19.65&-13.35&-17.77&-4.88&-0.68&0.56&-21.38&-19.73&-23.28                          \\
WITCHER3&-1.35&1.96&2.25&-6.68&-5.30&-8.52&-1.26&2.11&1.75&-6.45&-4.94&-8.31&-5.40&-8.35&-8.85&-15.54&-20.85&-21.71                                \\\midrule
\textbf{Average}&3.86&7.56&8.76&-2.93&0.30&-0.79&4.04&8.17&8.90&-2.75&0.58&-0.50&0.47&-0.51&0.16&-7.15&-10.21&-10.69                                        \\
\bottomrule
\end{tabular}}
\vspace{-4mm}
\end{table}

\begin{table}[!h]
\vspace{-2mm}
\setstretch{0.8}
\vspace{-0.5mm}
\caption{BD-Rate VMAF 4K NEG (\%) (Ada Lovelace HQ Baseline)}
\vspace{0.5mm}
\centering
\resizebox{17.5cm}{!}{\begin{tabular}{@{}lcccccccccccccccccc@{}}
\toprule
\multirow{2.5}{*}{Sequence Name}& \multicolumn{3}{c}{Ampere HEVC HQ} & \multicolumn{3}{c}{Ampere HEVC UHQ} & \multicolumn{3}{c}{Blackwell HEVC HQ} & \multicolumn{3}{c}{Blackwell HEVC UHQ}& \multicolumn{3}{c}{Blackwell AV1 HQ}&\multicolumn{3}{c}{Blackwell AV1 UHQ}\\
\cmidrule(lr){2-4} \cmidrule(lr){5-7} \cmidrule(lr){8-10} \cmidrule(lr){11-13} \cmidrule(lr){14-16} \cmidrule(lr){17-19}
& P1 & P4 & P7 & P1 & P4 & P7 & P1 & P4 & P7 & P1 & P4 & P7 & P1 & P4 & P7 & P1 & P4 & P7\\\midrule
\multicolumn{19}{c}{\textbf{Netflix Chimera}}\\\midrule
1\_BAR\_SCENE&-0.46&-0.12&-0.77&-14.01&-10.81&-13.01&-4.29&-6.94&-7.37&-14.51&-18.03&-19.74&-1.50&0.97&0.45&-25.24&-25.17&-25.32                  \\
2\_DINNER\_SCENE&-1.71&-0.97&-1.64&-29.06&-33.17&-39.13&-2.74&-5.34&29.45&-37.82&-42.28&-46.64&20.29&22.21&21.87&-31.99&-33.67&-33.21             \\
3\_DANCER&-3.86&-2.63&-3.04&-21.73&-22.26&-21.44&-6.11&-7.32&-7.70&-21.77&-24.36&-23.08&-0.64&1.05&-0.21&9.93&5.38&0.33                           \\
4\_DANCERS\_COUPLE&-1.25&-1.47&-1.22&-2.70&-9.28&-8.35&-3.77&-4.89&-4.59&-12.19&-17.71&-16.70&-2.59&-1.30&-1.50&-4.86&-6.23&-8.46                 \\
5\_DANCERS\_MONTAGE\_MIXED&-0.37&1.67&1.49&-19.14&-8.76&-9.60&-25.80&-2.66&-1.13&-22.60&-4.38&-3.36&9.58&13.12&13.02&2.23&1.30&-0.80              \\
6\_ROLLERCOASTER\_SEQUENCE&0.25&-0.16&-0.08&-1.80&-0.14&0.32&-0.24&-10.27&-10.26&-6.96&-14.30&-13.77&-5.69&-3.78&-3.70&-14.42&-14.17&-14.53       \\
7\_ROLLERCOASTER\_POV&0.08&-0.06&-0.12&-0.21&1.36&1.98&0.25&-8.66&-8.53&-4.63&-11.64&-11.33&-5.25&-3.25&-3.13&-14.63&-14.50&-14.58                \\
8\_ROLLERCOASTER\_PASSENGER&-0.37&-0.05&-0.14&-4.72&-3.19&-3.21&0.01&-9.96&-9.67&-5.56&-13.17&-13.16&-5.34&-3.58&-3.70&-16.07&-16.49&-16.85       \\
9\_TWIRL\_RIDE\_BOARDWALK&-0.26&-0.13&-0.06&1.15&2.70&2.85&-2.95&-10.78&-10.32&0.19&-4.50&-3.74&-6.53&-4.18&-3.81&-16.76&-15.74&-14.65            \\
10\_NETFLIX\_CARD\_TWIRL&0.09&0.00&0.03&-9.26&-9.68&-10.77&-3.40&-5.41&-5.54&-11.58&-15.59&-16.62&-5.59&-4.55&-4.45&-22.68&-23.61&-23.47          \\
11\_SEASIDE\_AND\_PIER&-0.77&-0.65&-0.55&-1.53&-0.68&-0.87&-4.27&-9.93&-9.94&0.03&-4.87&-5.41&-7.56&-5.53&-5.93&-16.54&-14.82&-14.70              \\
12\_WIND\_AND\_NATURE&-0.17&-0.22&-0.21&-8.40&-10.93&-14.42&-5.05&-7.23&-7.37&-28.50&-30.31&-32.84&-4.29&-3.63&-3.42&-29.10&-30.96&-30.93         \\
13\_MOUNTAIN\_VIEW\_W\_TILT&-2.48&-1.92&-2.22&-42.02&-37.50&-32.87&2.45&-24.95&-24.68&-41.90&-50.23&-52.05&-3.48&-2.33&-2.53&-50.18&-50.05&-49.48 \\
14\_MOUNTAIN\_VIEW\_PAN&-1.35&-0.24&0.39&-34.61&-21.13&-31.49&-0.26&-8.86&-7.90&-33.60&-36.03&-37.17&6.31&7.04&7.45&-44.07&-43.51&-45.51          \\
15\_WALK\_LIKE\_A\_MAN&-0.71&-0.87&-0.71&-15.77&-16.37&-17.29&-1.13&-10.61&-9.96&-11.24&-14.49&-16.39&-5.55&-4.69&-4.70&-26.35&-26.59&-35.40      \\
16\_TODDLER\_AND\_FOUNTAIN&-0.32&-0.21&-0.14&-0.63&-0.55&-1.41&-9.04&-10.80&-11.11&-14.01&-20.36&-20.66&-10.30&-9.48&-9.23&-13.28&-14.10&-11.01   \\
17\_DRIVING\_POV&-0.65&-0.55&-0.58&-0.69&0.14&-0.94&-2.02&-9.19&-8.99&1.13&-3.97&-4.53&-6.67&-5.26&-5.04&-17.93&-17.31&-17.04                     \\
18\_PLANET\_MOBILE&-0.76&-1.04&-0.71&-20.02&-21.69&-25.04&-1.27&1.88&2.73&-35.25&-50.83&-51.96&-11.02&-8.61&-8.88&-30.54&-32.08&-34.46            \\
19\_DOG\_PANTS&-0.57&-0.56&-1.41&-14.37&-14.43&-14.54&1.23&-12.70&-13.15&-11.50&-17.23&-19.37&-1.64&0.38&0.07&-20.02&-19.36&-20.68                \\
20\_DOG\_BARKS&-2.05&-2.70&0.28&-18.74&-21.86&-23.11&-3.28&-8.33&-6.74&-23.86&-28.20&-29.19&-3.63&-2.32&-2.31&-31.02&-32.15&-32.91                \\
21\_RC\_AERIAL&-1.33&-0.83&-0.63&-8.80&-7.70&-8.14&-0.62&-8.11&-8.02&-4.61&-8.02&-9.14&-6.87&-5.63&-5.48&-17.44&-16.67&-17.09                     \\
22\_BASKETBALL\_FREE\_THROW&-0.04&-0.74&-0.05&-4.17&-6.08&-5.91&-2.54&-10.36&-8.81&-15.28&-20.80&-20.96&-6.35&-4.95&-5.02&-17.01&-16.35&-17.99    \\
23\_BASKETBALL\_GAME&-0.43&0.38&0.31&2.84&4.65&5.38&-1.90&-8.47&-8.42&3.33&0.07&0.65&-6.26&-4.58&-4.70&-25.10&-13.38&-13.09                       \\\midrule
\textbf{Average}&-0.85&-0.61&-0.51&-11.67&-10.75&-11.78&-3.34&-8.69&-6.87&-15.33&-19.62&-20.31&-3.07&-1.43&-1.52&-20.57&-20.44&-21.38                      \\\midrule
\multicolumn{19}{c}{\textbf{Twitch}}\\\midrule
CSGO&-0.40&-0.56&-0.41&0.94&2.81&2.36&-3.42&-7.66&-7.09&-2.36&-6.13&-5.91&-6.99&-4.85&-4.54&-16.39&-15.05&-15.05                                  \\
DOTA2&0.04&-0.14&-0.25&-5.83&-1.11&-3.50&-4.74&-10.20&-11.29&-7.40&-12.74&-13.50&-8.01&-6.99&-7.21&-22.43&-22.86&-23.26                           \\
EuroTruckSimulator2&-0.91&-0.70&-0.63&-1.80&-10.03&-11.91&-5.15&-8.69&-8.75&-7.06&-13.97&-15.20&-8.11&-6.96&-6.73&-26.68&-28.12&-24.07            \\
Fallout4&-0.55&-0.41&-0.38&-6.54&-4.44&-4.66&-3.79&-7.34&-7.36&-7.71&-10.70&-11.01&-8.65&-7.11&-6.72&-18.72&-18.04&-18.33                         \\
GTAV&-0.60&-0.76&-0.61&-2.46&-2.04&-3.34&-4.03&-9.72&-10.13&-7.15&-13.90&-14.98&-8.19&-5.67&-5.85&-13.44&-23.17&-13.23                            \\
Hearthstone&0.50&-2.18&0.18&-26.65&-29.07&-31.52&-0.40&-5.78&-4.26&-28.08&-31.89&-34.46&-0.23&-4.62&1.56&-30.86&-38.33&-32.67                     \\
MINECRAFT&0.18&0.10&0.02&-3.48&-4.74&-5.97&-1.51&-5.75&-6.32&4.67&-5.23&-5.09&-4.63&-3.67&-1.62&-15.58&-11.72&-17.41                              \\
RUST&-0.42&-0.39&-0.31&2.80&1.29&-0.04&-6.21&-12.74&-12.80&-6.33&-15.69&-16.84&-10.85&-9.07&-9.06&-16.87&-16.76&-17.27                            \\
STARCRAFT&0.53&-0.34&0.71&-9.72&-9.82&-15.32&-0.49&-3.46&-3.21&-13.63&-19.02&-22.84&-6.55&-5.32&-5.14&-24.39&-24.80&-25.11                        \\
WITCHER3&-0.09&-0.15&0.49&-5.77&-8.11&-11.50&-4.36&-10.72&-10.99&-13.09&-21.28&-22.34&-13.13&-10.55&-10.62&-25.16&-25.92&-26.09                   \\\midrule
\textbf{Average}&-0.17&-0.55&-0.12&-5.85&-6.53&-8.54&-3.41&-8.21&-8.22&-8.81&-15.06&-16.22&-7.53&-6.48&-5.59&-21.05&-22.48&-21.25                          \\
\bottomrule
\end{tabular}}
\vspace{-4mm}
\end{table}

\begin{table}[!h]
\vspace{-2mm}
\setstretch{0.8}
\vspace{-0.5mm}
\caption{BD-Rate VMAF 4K NEG (\%) (HQ vs. UHQ)}
\vspace{0.5mm}
\centering
\resizebox{17.5cm}{!}{\begin{tabular}{@{}lcccccccccccccccccc@{}}
\toprule
\multirow{2.5}{*}{Sequence Name}& \multicolumn{3}{c}{Ampere HEVC} & \multicolumn{3}{c}{Ada Lovelace HEVC} & \multicolumn{3}{c}{Blackwell HEVC} & \multicolumn{3}{c}{Ada Lovelace AV1}& \multicolumn{3}{c}{Blackwell AV1}\\
\cmidrule(lr){2-4} \cmidrule(lr){5-7} \cmidrule(lr){8-10} \cmidrule(lr){11-13} \cmidrule(lr){14-16} 
& P1 & P4 & P7 & P1 & P4 & P7 & P1 & P4 & P7 & P1 & P4 & P7 & P1 & P4 & P7\\\midrule
\multicolumn{16}{c}{\textbf{Netflix Chimera}}\\\midrule
1\_BAR\_SCENE&-13.57&-10.70&-12.28&-14.11&-10.27&-12.35&-10.40&-11.71&-13.10&-20.85&-20.46&-20.62&-22.25&-23.11&-23.55               \\
2\_DINNER\_SCENE&-27.76&-32.51&-38.02&-28.74&-31.74&-37.83&-37.08&-40.75&-54.45&-28.59&-32.42&-33.36&-33.11&-34.39&-35.18            \\
3\_DANCER&-18.94&-20.61&-20.00&-17.67&-19.88&-19.49&-18.18&-19.01&-17.45&-1.22&-6.97&-9.69&10.62&3.86&0.37                           \\
4\_DANCERS\_COUPLE&-1.53&-8.03&-7.29&-0.53&-8.43&-7.12&-8.72&-13.28&-12.50&-1.99&-5.19&-6.15&-0.84&-3.66&-5.95                       \\
5\_DANCERS\_MONTAGE\_MIXED&-19.11&-10.59&-11.16&-18.33&-6.22&-9.10&4.81&-2.26&-2.58&-11.63&-14.95&-16.14&-5.29&-8.92&-10.87          \\
6\_ROLLERCOASTER\_SEQUENCE&-2.06&0.02&0.40&-1.64&-0.02&0.60&-6.73&-4.52&-4.15&-6.08&-7.49&-7.22&-9.04&-10.40&-10.96                  \\
7\_ROLLERCOASTER\_POV&-0.29&1.42&2.10&0.02&1.58&2.10&-4.88&-3.35&-3.33&-6.19&-7.65&-7.38&-9.69&-11.22&-11.47                         \\
8\_ROLLERCOASTER\_PASSENGER&-4.34&-3.15&-3.06&-4.59&-2.87&-2.83&-5.55&-4.02&-4.48&-7.83&-9.61&-9.52&-10.76&-12.73&-13.11             \\
9\_TWIRL\_RIDE\_BOARDWALK&1.42&2.84&2.91&1.24&2.77&2.94&2.63&5.17&5.44&-7.38&-8.62&-7.43&-10.38&-11.46&-10.79                        \\
10\_NETFLIX\_CARD\_TWIRL&-9.33&-9.66&-10.77&-9.21&-9.66&-10.83&-8.53&-10.84&-11.82&-14.86&-16.28&-15.98&-18.06&-19.86&-19.84         \\
11\_SEASIDE\_AND\_PIER&-0.76&-0.02&-0.31&-0.89&-0.03&-0.25&4.49&5.59&4.99&-8.12&-8.09&-8.31&-7.88&-8.01&-8.10                        \\
12\_WIND\_AND\_NATURE&-8.22&-10.72&-14.21&-8.23&-10.75&-14.21&-24.68&-24.81&-27.47&-15.33&-18.46&-18.73&-24.80&-27.30&-27.44         \\
13\_MOUNTAIN\_VIEW\_W\_TILT&-40.22&-36.22&-31.31&-41.17&-36.34&-31.17&-43.36&-34.09&-36.90&-43.21&-44.75&-44.38&-49.16&-49.71&-49.08 \\
14\_MOUNTAIN\_VIEW\_PAN&-33.70&-20.82&-31.65&-33.96&-19.93&-17.93&-32.68&-29.23&-30.81&-29.54&-29.01&-28.82&-46.92&-46.74&-48.54     \\
15\_WALK\_LIKE\_A\_MAN&-15.14&-15.62&-16.69&-15.14&-16.21&-16.89&-10.14&-4.32&-7.10&-19.60&-21.58&-22.07&-22.50&-23.45&-32.50        \\
16\_TODDLER\_AND\_FOUNTAIN&-0.32&-0.35&-1.27&-0.33&-0.29&-0.96&-5.42&-10.45&-10.53&-0.84&-1.55&-1.76&-3.18&-4.95&-1.90               \\
17\_DRIVING\_POV&-0.02&0.70&-0.35&-0.19&0.62&-0.49&3.20&5.25&4.39&-10.12&-10.27&-9.88&-11.72&-12.38&-12.31                           \\
18\_PLANET\_MOBILE&-19.39&-20.87&-24.49&-19.96&-21.09&-24.66&-34.11&-49.56&-50.95&-17.95&-18.82&-20.80&-17.56&-21.32&-23.98          \\
19\_DOG\_PANTS&-13.87&-13.93&-13.28&-14.37&-14.18&-14.79&-12.57&-5.10&-7.22&-16.94&-17.24&-17.61&-18.52&-18.95&-20.29                \\
20\_DOG\_BARKS&-16.70&-19.71&-23.18&-18.63&-21.54&-23.00&-21.16&-21.67&-24.03&-15.51&-16.74&-18.44&-28.65&-31.02&-31.85              \\
21\_RC\_AERIAL&-7.53&-6.90&-7.54&-7.94&-7.21&-7.59&-4.03&-0.28&-1.56&-10.66&-12.78&-13.27&-11.37&-11.54&-12.19                       \\
22\_BASKETBALL\_FREE\_THROW&-4.11&-5.37&-5.85&-3.85&-5.78&-5.45&-13.08&-11.77&-13.41&-7.73&-8.42&-9.05&-11.31&-12.11&-13.58          \\
23\_BASKETBALL\_GAME&3.28&4.27&5.06&2.82&4.63&5.21&5.16&9.04&9.54&-6.53&-7.28&-6.27&-19.78&-8.85&-8.48                               \\\midrule
\textbf{Average}&-10.97&-10.28&-11.40&-11.10&-10.12&-10.70&-12.22&-12.00&-13.46&-13.42&-14.98&-15.34&-16.62&-17.75&-18.76                     \\\midrule
\multicolumn{16}{c}{\textbf{Twitch}}\\\midrule
CSGO&-2.29&-2.28&-3.23&-2.42&-2.59&-3.31&-3.52&-4.07&-4.67&-7.42&-7.87&-8.06&-7.55&-7.92&-8.40                                       \\
DOTA2&-5.30&-0.88&-2.15&-5.23&-0.93&-2.21&-2.94&-4.36&-3.74&-15.15&-15.56&-16.00&-13.94&-15.04&-15.75                                \\
EuroTruckSimulator2&-2.57&-3.29&-4.78&-2.88&-3.22&-4.79&-2.53&-5.50&-6.63&-13.31&-14.87&-15.14&-13.27&-15.50&-16.31                  \\
Fallout4&-0.59&0.41&0.32&-0.69&0.42&0.38&0.10&-0.38&-0.50&-4.93&-5.97&-5.98&-3.70&-4.77&-5.35                                        \\
GTAV&-0.89&0.67&-0.34&-1.09&0.29&-0.45&-1.72&-2.24&-2.52&-4.46&-4.73&-5.73&-5.32&-7.74&-7.40                                         \\
Hearthstone&-33.20&-33.64&-37.73&-33.07&-35.52&-37.28&-33.02&-32.80&-36.90&-12.95&-21.57&-15.77&-18.60&-20.64&-22.04                 \\
MINECRAFT&-3.48&-7.08&-7.80&-3.69&-6.82&-7.74&-4.64&-10.36&-9.61&-9.12&-10.95&-12.00&-5.84&-8.21&-9.34                               \\
RUST&2.99&1.67&0.31&2.86&1.60&0.32&0.53&-2.00&-3.08&-3.91&-5.90&-6.67&-4.37&-6.26&-7.09                                              \\
STARCRAFT&-16.27&-14.97&-20.75&-15.79&-15.25&-20.52&-17.04&-19.07&-23.60&-14.36&-15.18&-17.32&-18.17&-19.01&-20.14                   \\
WITCHER3&-5.50&-7.18&-10.59&-5.33&-6.97&-9.94&-10.67&-13.73&-14.16&-9.13&-11.50&-13.64&-11.35&-14.00&-14.46                          \\\midrule
\textbf{Average}&-6.71&-6.66&-8.67&-6.73&-6.90&-8.55&-7.55&-9.45&-10.54&-9.47&-11.41&-11.63&-10.21&-11.91&-12.63                              \\
\bottomrule
\end{tabular}}
\vspace{-4mm}
\end{table}

\begin{table}[!h]
\vspace{-2mm}
\setstretch{0.8}
\vspace{-0.5mm}
\caption{BD-Rate VMAF 4K NEG (\%) (Ada Lovelace UHQ Baseline)}
\vspace{0.5mm}
\centering
\resizebox{12cm}{!}{\begin{tabular}{@{}lcccccccccccccccccc@{}}
\toprule
\multirow{2.5}{*}{Sequence Name}& \multicolumn{3}{c}{Ampere HEVC UHQ} & \multicolumn{3}{c}{Blackwell HEVC UHQ} & \multicolumn{3}{c}{Blackwell AV1 UHQ} \\
\cmidrule(lr){2-4} \cmidrule(lr){5-7} \cmidrule(lr){8-10} 
& P1 & P4 & P7 & P1 & P4 & P7 & P1 & P4 & P7 \\\midrule
\multicolumn{10}{c}{\textbf{Netflix Chimera}}\\\midrule
1\_BAR\_SCENE&0.27&-0.25&-0.27&-9.27&-17.80&-15.58&1.17&-2.06&-1.31                  \\
2\_DINNER\_SCENE&-0.29&-0.98&-1.27&-21.98&-22.87&-19.28&6.10&9.57&12.33              \\
3\_DANCER&-2.48&-1.89&-1.75&-32.60&-29.60&-24.76&14.48&14.09&11.50                   \\
4\_DANCERS\_COUPLE&-1.43&-0.37&-0.78&-15.66&-15.84&-14.76&7.30&9.98&8.86             \\
5\_DANCERS\_MONTAGE\_MIXED&-0.71&-2.13&-0.15&-24.19&-22.03&-17.41&19.84&24.42&22.68  \\
6\_ROLLERCOASTER\_SEQUENCE&0.09&-0.06&-0.31&4.15&-7.88&-7.08&-5.23&-3.54&-4.51       \\
7\_ROLLERCOASTER\_POV&-0.19&-0.22&-0.01&4.22&-6.70&-5.91&-5.09&-3.34&-4.35           \\
8\_ROLLERCOASTER\_PASSENGER&-0.01&-0.26&-0.28&-0.31&-12.52&-11.88&-5.37&-3.65&-4.57  \\
9\_TWIRL\_RIDE\_BOARDWALK&0.04&0.06&-0.08&-5.60&-13.49&-12.97&-7.93&-5.15&-5.41      \\
10\_NETFLIX\_CARD\_TWIRL&-0.12&-0.02&0.16&-7.19&-9.23&-8.95&-6.05&-6.15&-6.51        \\
11\_SEASIDE\_AND\_PIER&-0.48&-0.44&-0.38&-7.45&-14.10&-13.84&-7.93&-6.22&-6.19       \\
12\_WIND\_AND\_NATURE&-0.06&-0.12&-0.19&15.56&12.07&12.82&1.22&1.96&2.61             \\
13\_MOUNTAIN\_VIEW\_W\_TILT&0.03&-1.76&-2.16&0.93&-20.32&-27.28&4.53&14.32&12.51     \\
14\_MOUNTAIN\_VIEW\_PAN&-0.07&0.65&-29.49&-7.63&-36.41&-36.77&7.35&8.14&6.36         \\
15\_WALK\_LIKE\_A\_MAN&-0.48&-0.27&-0.16&-5.53&-10.46&-10.97&3.12&-4.52&6.62         \\
16\_TODDLER\_AND\_FOUNTAIN&1.35&-0.46&-0.45&-9.99&-11.27&-11.52&-10.81&-9.56&-9.39   \\
17\_DRIVING\_POV&-0.39&-0.35&-0.30&-4.31&-11.43&-10.66&-6.72&-5.28&-5.68             \\
18\_PLANET\_MOBILE&0.08&-0.69&-0.54&-12.04&-27.62&-25.08&-3.00&-3.80&-4.73           \\
19\_DOG\_PANTS&0.20&-0.06&0.64&0.22&-6.23&-7.48&0.29&0.78&-0.46                      \\
20\_DOG\_BARKS&-0.05&0.13&-0.58&-3.02&-6.16&-4.43&4.08&5.36&4.90                     \\
21\_RC\_AERIAL&-0.67&-0.15&-0.24&0.49&-7.29&-7.86&5.14&-7.14&-7.48                   \\
22\_BASKETBALL\_FREE\_THROW&-0.19&-0.39&0.14&-3.55&-8.77&-8.25&-3.80&-1.35&-0.78     \\
23\_BASKETBALL\_GAME&0.23&-9.38&-0.07&5.64&-10.08&-9.26&-6.34&-5.58&-6.55            \\\midrule
\textbf{Average}&-0.23&-0.84&-1.67&-6.05&-13.74&-13.01&0.28&0.93&0.89                         \\\midrule
\multicolumn{10}{c}{\textbf{Twitch}}\\\midrule
CSGO&3.50&5.70&5.85&-4.65&-9.04&-8.35&-7.83&-5.77&-5.74                              \\
DOTA2&-0.62&-0.17&-1.31&-2.54&-13.05&-12.39&-7.07&-6.80&-6.88                        \\
EuroTruckSimulator2&1.12&-6.96&-7.31&-5.05&-10.93&-10.62&-8.44&-7.98&-8.17           \\
Fallout4&-5.86&-4.76&-5.00&-3.21&-8.17&-8.27&-8.01&-6.58&-6.66                       \\
GTAV&-1.31&-2.32&-2.90&-4.79&-12.18&-12.17&-10.10&-10.28&-8.91                       \\
Hearthstone&9.50&9.88&9.20&1.27&0.53&-0.44&-5.62&-5.11&-4.78                         \\
MINECRAFT&0.21&2.25&1.94&-2.65&-9.51&-8.35&-2.09&-0.79&-0.52                         \\
RUST&-0.06&-0.28&-0.37&-8.42&-15.87&-15.79&-12.13&-10.31&-10.12                      \\
STARCRAFT&7.20&6.39&6.50&-1.54&-6.89&-6.05&-9.66&-8.71&-7.54                         \\
WITCHER3&-0.45&-1.23&-1.68&-9.70&-16.24&-13.67&-17.25&-15.75&-13.68                  \\\midrule
\textbf{Average}&1.32&0.85&0.49&-4.13&-10.14&-9.61&-8.82&-7.81&-7.30                          \\
\bottomrule
\end{tabular}}
\vspace{-4mm}
\end{table}

\end{document}